\documentclass[acmsmall]{acmart}

\usepackage{xspace}
\usepackage{xcolor}
\usepackage{algorithm}
\usepackage{algpseudocode}
\usepackage{listings}
\AtBeginDocument{%
  }

\newcommand{\sys}{\mbox{\textsc{SBridge}}\xspace}

\acmConference[FSE '26]{
the ACM International Conference on the Foundations of Software Engineering}{July 5--9, 2026}{ Montreal, Canada}

\usepackage{xstring}
\newcommand{\PP}[1]{
	\vspace{4px}
{\noindent\bf\textit{\IfEndWith{#1}{.}{#1}{#1.}}}}

\newcommand{\ie}{\textit{i}.\textit{e}.}
\newcommand{\eg}{\textit{e}.\textit{g}.}

\usepackage{caption}
\usepackage{colortbl}
\usepackage{arydshln}
\usepackage{kotex}
\usepackage{xcolor}
\usepackage{url}
\usepackage{soul}
\PassOptionsToPackage{hyphens}{url}
\usepackage{hyperref}
\hypersetup{
	colorlinks,
	linkcolor={red!100!black},
	citecolor={red!100!black},
	urlcolor={icseorange!100!black}
}
\definecolor{citegray}{rgb}{0.4, 0.4, 0.4}
\definecolor{tomato}{rgb}{0.7, 0.0, 0.0}
\definecolor{mypink}{rgb}{0.98, 0.925, 0.929}
\definecolor{myorange}{rgb}{0.988235294,0.850980392,0.733333333}
\definecolor{myyellow}{rgb}{0.9647058823529412, 0.9686274509803922, 0.7686274509803922}
\definecolor{mydiffgreen}{rgb}{0.855, 0.96, 0.886}
\definecolor{myblue}{rgb}{0.807843137, 0.984313725, 0.964705882}
\definecolor{myindigo}{rgb}{0.796078431,0.941176471,1}
\definecolor{mypurple}{rgb}{0.929411765,0.941176471,1}

\definecolor{blue1}{RGB}{191,214,135}
\definecolor{blue2}{rgb}{0.8,0.8,0.8}
\definecolor{blue3}{RGB}{255, 236, 139} 
\definecolor{blue4}{rgb}{0.6,0.6,0.6}
\definecolor{blue5}{rgb}{0.7,0.7,0.7}
\definecolor{blue6}{rgb}{0.8,0.8,0.8}
\definecolor{blue7}{rgb}{0.9,0.9,0.9}

\definecolor{fsemint}{RGB}{141,211,199}        
\definecolor{fseorange}{RGB}{217, 242, 208}
\definecolor{fsepurple}{RGB}{255, 255, 213} 
\definecolor{fsepink}{RGB}{251,128,114}    

\usepackage[skins]{tcolorbox}
\tcbuselibrary{breakable}
\usepackage{kotex}
\usepackage{subcaption}
\usepackage{listings}
\usepackage{multirow}
\usepackage{breakurl}
\usepackage{amsmath}

\def\sectionautorefname~#1\null{Section #1\null}
\def\equationautorefname~#1\null{Equation (#1)\null}
\def\algorithmautorefname~#1\null{Algorithm (#1)\null}
\def\subsectionautorefname~#1\null{Section #1\null}
\def\subsubsectionautorefname~#1\null{Section #1\null}

\lstdefinestyle{base}{
	language=C,
	numbers=left,
	numberstyle=\tiny,
	numbersep=-3pt,
	breaklines=true,
	commentstyle=\color{mygray},
	columns=fullflexible,
	basicstyle=\fontsize{7.3}{9}\ttfamily\color{black},
	moredelim=**[is][\bfseries]{@bold-}{-bold@},
	moredelim=**[is][\color{tomato}]{@R-}{-R@},
	moredelim=**[is][\color{mygreen}]{@G-}{-G@},
	moredelim=**[is][\color{mygray}\bfseries]{@B-}{-B@},
	moredelim=**[is][\color{blue}\bfseries]{@AB-}{-AB@},
	escapeinside={(**}{**)},
	tabsize=4,
	xleftmargin=1pt,
	captionpos=t,
	aboveskip=0.65\baselineskip,
	belowskip=0.65\baselineskip,
	showstringspaces=false
}

\usepackage{float}

\setcopyright{cc}
\setcctype{by}
\acmDOI{10.1145/3797090}
\acmYear{2026}
\acmJournal{PACMSE}
\acmVolume{3}
\acmNumber{FSE}
\acmArticle{FSE062}
\acmMonth{7}
\acmSubmissionID{fse26maina-p1590-p}
\received{2025-09-12}
\received[accepted]{2025-12-22}

\begin{document}

\title{\sys: Identifying Source-to-Binary Function Similarity via Cross-Domain Control Block Matching}


\author{Heedong Yang}
\orcid{0000-0003-3121-8064}
\affiliation{%
  \institution{Korea University}
  \city{Seoul}
  \country{Republic of Korea}
}
\email{heedongy@korea.ac.kr}

\author{Jeongwoo Lee}
\orcid{0009-0004-9901-2673}
\affiliation{%
  \institution{Korea University}
  \city{Seoul}
  \country{Republic of Korea}
}
\email{jeongwoo@korea.ac.kr}

\author{Hajin Yun}
\orcid{0009-0006-5345-8447}
\affiliation{%
  \institution{Korea University}
  \city{Seoul}
  \country{Republic of Korea}
}
\email{hajinyun1011@korea.ac.kr}

\author{Seunghoon Woo}
\authornote{Corresponding author}
\orcid{0000-0002-5455-0804}
\affiliation{%
  \institution{Korea University}
  \city{Seoul}
  \country{Republic of Korea}
}
\email{seunghoonwoo@korea.ac.kr}

\begin{abstract}
    We present \sys, a precise approach for identifying functions in binaries that are similar to the given source code functions. Identifying reused code in binaries is critical for security, particularly for detecting propagated vulnerabilities.Although binary-to-binary comparison is feasible, leveraging source code as the reference is more practical because source code is easier to collect and analyze directly without compilation.However, significant gaps between source and binary representations, including function inlining, create challenges in cross-domain function detection.Existing approaches primarily rely on string literals or structural similarities between entire functions, failing to capture detailed code behavior and generating many false alarms. 
    
    \sys addresses these limitations through a key innovation: control block-based function matching, which encapsulates essential functional features by segmenting functions into meaningful units such as conditionals and loops. Leveraging control blocks as a cross-domain representation, \sys enables precise measurement of function similarity between source and binary code, effectively overcoming challenges posed by function inlining and stripped binaries.
    For evaluation, we collected 3,904 real-world C/C++ binaries from BinKit. In experiments identifying binary functions identical to input source functions, despite approximately 40\% of binary functions being inlined, \sys achieved 75.13\% \textit{recall@1} and 80.98\% \textit{recall@5}, outperforming existing approaches, which achieved up to 43.31\% \textit{recall@1} and 50.2\% \textit{recall@5}. Our further analysis confirmed that \sys effectively identifies propagated vulnerabilities in binaries.
\end{abstract}

\ccsdesc[500]{Software and its engineering~Software maintenance tools}
\ccsdesc[300]{Security and privacy~Software security engineering}

\keywords{Source-to-Binary Matching, Clone Detection, Vulnerability Management.}


\maketitle

\section{Introduction}


Open-source software (OSS) has become ubiquitous in software development~\cite{blackduck2025report, woo2023v1scan}.
However,
this widespread adoption has introduced new challenges in software security~\cite{kim2017vuddy, xiao2020mvp}. 
In particular, as Commercial-off-the-shelf (COTS) binaries increasingly incorporate OSS components, 
reused code analysis has become essential for vulnerability prevention and license compliance~\cite{blackduck2025report,duan2017identifying,woo2025large}.

Therefore,
\textit{code similarity measurement} has emerged as a key technique for
enabling 
software composition analysis (SCA) 
to manage 
OSS components in COTS binaries~\cite{yuan2019b2sfinder,woo2021centris,dong2024libvdiff,na2024cneps,jiang2023third} and detect propagated vulnerabilities~\cite{woo2022movery,woo2021v0finder,woo2023v1scan,zhan2024react}.
%
Traditionally, \textit{binary-to-binary} matching is dominated by comparing target binaries against binary code databases
(\eg, \cite{yang2022modx,tang2020libdx,li2023libam,xu2023pem}). However, this approach is impractical for real-world use because maintaining a database of numerous compiled software variants is costly and time-consuming. 

To address this issue, \textit{source-to-binary} matching emerged (\eg, \cite{duan2017identifying, yuan2019b2sfinder, jiang2024binaryai}), leveraging source code databases and focusing on compilation-resilient features (\eg, string literals).

Although source-to-binary matching effectively identifies OSS,
existing efforts typically operate at a coarse granularity, lacking precision for detailed function-level analysis.
A more refined approach could enable precise detection of binary code fragments (\eg, functions) that closely correspond to source inputs, thereby
enhancing both binary SCA and vulnerability detection.



However, achieving precise source-to-binary matching faces the following three fundamental challenges (details are introduced in \autoref{subsec:problem}).



%
\vspace{-0.4em}
\begin{enumerate}
    \setlength\itemsep{0.3em}
    \item \textbf{Loss of source code context.} 
    The compilation process strips crucial information (\eg, variable names), widening the gap between source and binary representations. 
    %


    \item \textbf{Architecture and compiler variability.} 
    Binary code structure and behavior vary significantly across different hardware architectures and compiler configurations. 
   
    
    %

    \item \textbf{Addressing function inlining.}
    Modern compilers optimize performance by inlining functions, embedding callee functions' code at call sites. 

    
\end{enumerate}

\vspace{-4px}
\PP{Limitations of existing approaches}
Existing source-to-binary matching approaches (\eg, \cite{duan2017identifying, yuan2019b2sfinder, jiang2024binaryai}) have focused on coarse-grained OSS detection. 
Therefore, they are ineffective for function similarity measurement, especially due to the aforementioned challenges.
For example,
B2SFinder~\cite{yuan2019b2sfinder} relies on features susceptible to compiler-induced variations (\eg, control flow), reducing efficiency in source-binary matching. Meanwhile, the state-of-the-art approach BinaryAI~\cite{jiang2024binaryai} overlooks function inlining, leading to low accuracy in identifying binary functions similar to source functions (see \autoref{subsec:rq1}).
Alternatively, a source-to-source matching strategy (\eg, \cite{sajnani2016sourcerercc, wang2018ccaligner, yu2025multiple}) can also be considered, where \textit{decompiled} binaries are compared with source code. However, decompiled code often deviates significantly from the original source in syntax, leading to low detection accuracy. Moreover, such approaches are particularly ineffective when function inlining has occurred. 

\vspace{4px}
To overcome their shortcomings, we present \sys (\underline{S}ource-to-\underline{B}inary b\underline{\textsc{ridge}}), a precise approach for identifying functions in binaries that are similar to the input source functions.

\PP{Our approach.}
The key technical contribution of \sys is a \textit{control block-based function similarity measurement} technique, which introduces a new granularity called control blocks to capture essential structural and semantic features for accurate source-to-binary matching.


Given input source functions and target binary, \sys first extracts \textit{control blocks}—fine-grained units encapsulating structural and semantic features (\eg, \texttt{if} condition blocks)—from both the source and decompiled binary functions (see \autoref{subsec:p1}).
The core insight is that control blocks preserve their functional integrity 
even after function inlining, 
enabling \sys to identify cross-domain function similarities.
\sys considers both internal and external branching patterns of functions, and categorizes control blocks into seven types
(see \autoref{table:controlblock}).
It then extracts \textit{key features} essential for matching (\eg, condition expressions) and \textit{block contents} used as auxiliary information (\eg, syntax of code lines within conditional statements) 
from each block (see \autoref{subsec:p1}).

\sys computes function similarity based on the similarity of their constituent blocks. Using the source function's blocks as reference points, it quantifies the ratio of these blocks contained within a binary function, thereby preserving block-level similarity even when multiple source functions are inlined into a binary function. In addition, a weighting mechanism based on call relationships and function length is applied to further address function inlining (see \autoref{subsec:p2}). Block similarity is computed from key features and block content, and function similarity is finally derived from the proportion of matching blocks between two functions (see \autoref{subsec:p3}). 
Consequently, \sys can identify, for each input source function, the corresponding similar functions within the target binary.

\PP{Evaluation}
For our experiments, we collected 3,904 real-world C/C++ binaries from BinKit. Using the source code of each binary as input, we evaluated the ability to locate compiled functions from their source counterparts.
When compiling with \texttt{O2} optimization, we observed that approximately 40\% of source functions underwent function inlining during compilation  (see \autoref{subsec:rq2}).
Nevertheless, \sys achieved 75.13\% \texttt{recall@1} and 80.98\% \texttt{recall@5}, significantly outperforming existing approaches~\cite{jiang2024binaryai, yu2025multiple} (up to 43.31\% \texttt{recall@1} and 50.2\% \texttt{recall@5}; see \autoref{subsec:rq1}).
Further experiments on \sys's effectiveness and performance demonstrated that \sys (1) robustly maps source and binary functions even in the presence of function inlining (see \autoref{subsec:rq2}), (2) detects functions within an average of 1.9 s per binary, proving its practicality (see \autoref{subsec:rq3}), and (3) effectively identifies propagated vulnerabilities in binaries (see \autoref{subsec:rq4}).



\PP{Contributions} This paper makes the following three contributions.

\begin{itemize}
    \setlength\itemsep{0.2em}
    \item We present \sys, an approach to effectively detect binary functions similar to given source functions. Its key contribution is a control block-based function matching technique.
        
    \item \sys outperforms existing approaches by achieving a \texttt{recall@1} of 75.13\% and a \texttt{recall@5} of 80.98\% in mapping source functions to their compiled binary counterparts. 

    \item By introducing an effective source-to-binary function mapping method, \sys establishes a foundation for future advancements in static analysis of source and binary code. 

\end{itemize}

\section{Motivation}
In this section, we clarify the target problem addressed by \sys and explore its motivation through a motivating example.

\subsection{Problem and Technical Challenges}\label{subsec:problem}
We focus on the problem of precisely detecting similar code by comparing source code and binary. 
Specifically,
given a set of source code functions, 
\sys attempts to identify and map a set of binary functions 
within a target binary that exhibit high similarity to the input source functions. 

\begin{figure}
    \centering
    \includegraphics[width=0.9\linewidth]{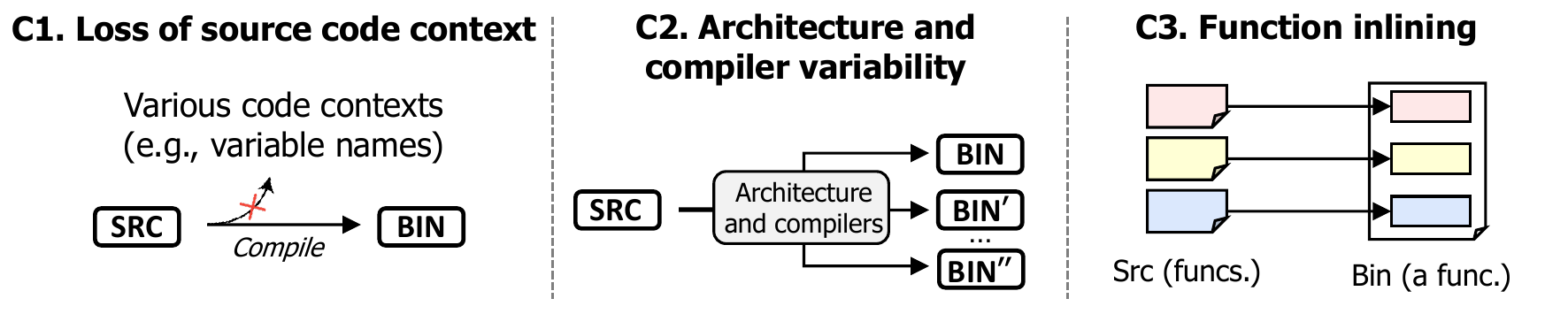}

    \vspace{-0.8em}
    \caption{Three major challenges in matching source code and binaries.}
    \label{fig:challenge}

\end{figure}

However, this \textit{source-to-binary} matching problem is particularly complex owing to the transformative nature of the compilation process and the diversity of execution environments.
Hence, to effectively perform source-to-binary matching, the following three technical challenges should be addressed (see \autoref{fig:challenge}).





\PP{C1: Loss of source code context.}
Many details contained in the source code are lost during compilation into binaries or when converting it into more human-readable low-level pseudocode (\eg, decompiled code). For example, variable names, data types, and the structure of the code are not preserved in the same way from source code to binaries. In particular, the distribution of most COTS software is as stripped binaries with static symbol tables (\eg, \texttt{.symtab} and \texttt{.strtab} sections) removed for security or licensing reasons. 
This semantic gap presents a fundamental challenge in making meaningful comparisons 
between source and binary functions.


\PP{C2: Architecture and compiler variability.}
Different operating systems and processor architectures  significantly impact how code is executed and represented in binary,
including variations in system calls and library implementations.
Compiler selection, versions, and optimization settings further change the compiled output, making it difficult to match binaries to their original source. Even with the same source function, different compilers may use different standard C/C++ libraries based on optimization options. Addressing this diversity in architecture and compilers remains a key challenge.
The diversity in architecture and compilers also represents a significant problem that needs to be addressed.

\PP{C3: Addressing function inlining.} 
Function inlining poses a major challenge as it follows complex, unpredictable patterns, often applied for performance optimization or reducing function call overhead. In binaries, multiple source functions may be merged into a single binary function (see \autoref{subsec:moti}), complicating similarity analysis between source and binary functions. Despite existing efforts to tackle function inlining, achieving a robust mapping from source to binary functions continues to be a significant challenge~\cite{jia20231,jia2024codeextract,jia2024cross}.

\subsection{Motivating Example}\label{subsec:moti}
To motivate our goal,
we examined the \texttt{csplit} binary from GNU Coreutils 9.0,
especially focusing on 
the \texttt{main}, \texttt{parse\_patterns}, and \texttt{extract\_regexp} functions. 
The binary was compiled for the ARM 32-bit architecture using Clang 13.0 with \texttt{-O2} optimization and stripped of symbols. 
This example highlights how the selected challenges, particularly function inlining, manifest in practice.


\autoref{lst:csplit_main}, \autoref{lst:csplit_parse_patterns}, and \autoref{lst:csplit_extract_regex} present source code snippets of three selected functions, while \autoref{lst:moti_decom_ex} shows a decompiled code snippet of the \texttt{main} function from the \texttt{csplit} binary.

A key observation is \textit{function inlining}: in the source code, 
\texttt{main} calls \texttt{parse\_patterns}, which in turn calls \texttt{extract\_regexp}. However, both the \texttt{parse\_patterns} and \texttt{extract\_regexp} functions are inlined into \texttt{main} in the target binary. 
Moreover, compilation causes substantial information loss; for example, variable and parameter names are removed and altered, literal values are modified (\eg, \textcolor{blue}{\texttt{`f'}} $\rightarrow$ \textcolor{blue}{\texttt{0x66}}), and function call statements disappear due to inlining (\eg, line \#8 in \autoref{lst:csplit_main}).

\PP{Existing approaches}
Function inlining, together with information loss during compilation, significantly undermines the effectiveness of existing source-to-binary matching approaches
(\eg, \cite{yuan2019b2sfinder, jiang2024binaryai}). 
For example,
they would attempt to find a source function corresponding to \texttt{FUN\_000115f4}, but may only match one of the three original functions or fail to identify a corresponding source function. Similarly, source-to-binary matching becomes challenging, as the syntax of \texttt{FUN\_000115f4} differs significantly from the original source functions.
These tendencies were most prominently observed in the evaluation, manifesting as limitations of existing techniques (see \autoref{subsec:rq1}).

\begin{figure}[t]
\begin{subfigure}[t]{0.32\linewidth}
\begin{lstlisting}[caption={The \texttt{main} function of \texttt{csplit}.\label{lst:csplit_main}}, language=C, style=base, basicstyle=\fontsize{5.6}{6.2}\ttfamily\color{black}, deletekeywords={for}, mathescape, breakatwhitespace=true, breaklines=true]
$\text{\quad\textcolor{black}{int \textbf{main} (int argc, char **argv)}}$
$\text{\quad\textcolor{black}{\{ ...}}$
$\text{\quad\;\;\;\textcolor{black}{switch (optc) \{}}$
$\text{\quad\;\;\;\;\;\;\textcolor{black}{case `\textcolor{blue}{f}':}}$
$\text{\quad\;\;\;\;\;\;\;\;\;\textcolor{black}{prefix = optarg;}}$
$\text{\quad\;\;\;\;\;\;\;\;\;\textcolor{black}{break;}}$
$\text{\quad\;\;\;\textcolor{black}{...}}$
$\text{\quad\;\;\;\textcolor{black}{\textbf{parse\_patterns} (}}$
$\text{\quad\;\;\;\;\;\;\textcolor{black}{argc, optind, argv);}}$
\end{lstlisting}
\end{subfigure}\hfill
\begin{subfigure}[t]{0.32\linewidth}

\begin{lstlisting}[caption={The \texttt{parse\_patterns} function of \texttt{csplit}.\label{lst:csplit_parse_patterns}}, language=C, style=base, basicstyle=\fontsize{5.6}{6.2}\ttfamily\color{black}, deletekeywords={for}, mathescape, breakatwhitespace=true, breaklines=true]
$\text{\quad\textcolor{black}{static void \textbf{parse\_patterns} (}}$
$\text{\quad\;\;\;\textcolor{black}{int argc, int start, char *argv) \{}}$
$\text{\quad\;\;\;\textcolor{black}{...}}$
$\text{\quad\;\;\;\textcolor{black}{if \textcolor{tomato}{(*argv[i] == `/' ||}}}$
$\text{\quad\;\;\;\;\;\;\textcolor{tomato}{*argv[i] == `\%')}}$
$\text{\quad\;\;\;\;\;\;\textcolor{black}{p = \textbf{extract\_regexp} (i,}}$
$\text{\quad\;\;\;\;\;\;\textcolor{black}{*argv[i] == `\%', argv[i]);}}$
$\text{\quad\;\;\;\textcolor{black}{...}}$
\end{lstlisting}
\end{subfigure}\hfill
\begin{subfigure}[t]{0.32\linewidth}

\begin{lstlisting}[caption={The \texttt{extract\_regexp} function of \texttt{csplit}.\label{lst:csplit_extract_regex}}, language=C, style=base, basicstyle=\fontsize{5.6}{6.2}\ttfamily\color{black}, deletekeywords={for}, mathescape, breakatwhitespace=true, breaklines=true]
$\text{\quad\textcolor{black}{static struct control *}}$
$\text{\quad\textcolor{black}{\textbf{extract\_regexp} (int argnum,}}$
$\text{\quad\;\;\;\textcolor{black}{bool ignore, char const *str)}}$
$\text{\quad\textcolor{black}{\{ ...}}$
$\text{\quad\;\;\;\textcolor{black}{char delim = *str;}}$
$\text{\quad\;\;\;\textcolor{black}{char const *closing\_delim;}}$
$\text{\quad\;\;\;\textcolor{black}{closing\_delim =}}$
$\text{\quad\;\;\;\;\;\;\textcolor{ACMGreen}{strrchr (str + 1, delim);}}$
$\text{\quad\;\;\;\textcolor{black}{if \textcolor{ACMPurple}{(closing\_delim == NULL)}}}$
$\text{\quad\;\;\;\textcolor{black}{...}}$
\end{lstlisting}
\end{subfigure}

\begin{lstlisting}[caption={Decompiled code of \texttt{main} (\texttt{FUN\_000115f4} in the stripped binary) extracted from the \texttt{csplit} binary.\label{lst:moti_decom_ex}}, language=C, style=base, basicstyle=\fontsize{7}{8}\ttfamily\color{black}, deletekeywords={for}, mathescape, breakatwhitespace=true, breaklines=true]
$\text{\quad\textcolor{black}{undefined4 \textbf{FUN\_000115f4} (int param\_1, undefined4 *param\_2) \{}}$
$\text{\quad\;\;\;\textcolor{black}{...}}$
$\text{\quad\;\;\;\textcolor{black}{switch(iVar7) \{}}$
$\text{\quad\;\;\;\;\;\;\textcolor{black}{case \textcolor{blue}{0x66}:}}$
$\text{\quad\;\;\;\;\;\;\;\;\;\textcolor{black}{*piVar17 = *DAT\_00012744;}}$
$\text{\quad\;\;\;\;\;\;\;\;\;\textcolor{black}{break;}}$
$\text{\quad\;\;\;\textcolor{black}{...}}$
$\text{\quad\;\;\;\textcolor{black}{\textcolor{tomato}{if (uVar28 == 0x2f || uVar28 == 0x25)} \{}}$
$\text{\quad\;\;\;\;\;\;\textcolor{black}{pcVar11 = \textcolor{ACMGreen}{strrchr((char *)(pbVar24 + 1), uVar28);}}}$
$\text{\quad\;\;\;\;\;\;\textcolor{black}{if \textcolor{ACMPurple}{(pcVar11 == (char *)0x0)}}}$
$\text{\quad\;\;\;\textcolor{black}{...}}$
\end{lstlisting}
\Description{Source snippets of the \texttt{main}, \texttt{parse\_patterns}, and \texttt{extract\_regexp} functions, followed by a decompiled binary snippet of \texttt{FUN\_000115f4}.}
\end{figure}

\vspace{4px}
\noindent\textit{\textbf{\sys.}}
\sys aims to identify binary functions similar to input source functions, particularly in cases where decompiled code deviates significantly from its source. The approach centers on extracting control blocks from each function. For example, the \texttt{switch-case} statement in \autoref{lst:csplit_main} is extracted as a \texttt{Condition} block, while the call to \texttt{parse\_patterns} is identified as a \texttt{CalleeFunction} block (see \autoref{subsec:p1}).
\sys's block matching approach, designed to recognize code changes in block structures, demonstrates that even when the original  \texttt{CalleeFunction} block from the source \texttt{main} function is not found in the decompiled code, the presence of various blocks (including \texttt{Condition} blocks and those omitted in \autoref{lst:csplit_main}) within the decompiled function indicates that \texttt{main} maintains high similarity with its decompiled counterpart. Similarly, \sys identifies that the remaining two functions also exhibit high similarity to the \texttt{FUN\_000115f4} function.
Notably, \sys's weighting mechanism based on call relationships and function length helps better account for inlining effects during similarity measurement (see \autoref{subsec:p3}).

\section{Design of \sys}\label{sec:design}

In this section, we describe the design of \sys, an approach for precisely detecting functions in a target binary that are similar to given source functions.

\PP{Overview}
\sys utilizes a new unit called a \textit{control block} to identify binary functions similar to source functions (see \autoref{subsec:p1}). The brief definition of a control block is as follows. 

\vspace{0.4em}
    \begin{center}
    \small
    \setlength{\fboxsep}{0.4em}
    \noindent\fbox{%
    \parbox{0.95\linewidth}{%
    	{{\scriptsize$\bullet$} \underline{\textbf{Definition: Control block}}
        \vspace{0.3em}

        \hspace{3px} A control block is a fundamental code unit within a function that encapsulates key functional features and represents distinct structural components, such as loops, if-else blocks, and embedded string elements.
        }%
    }
    }
    \end{center}
    \vspace{0.4em}



\begin{figure}[t]
    \centering
    \includegraphics[width=1\textwidth]{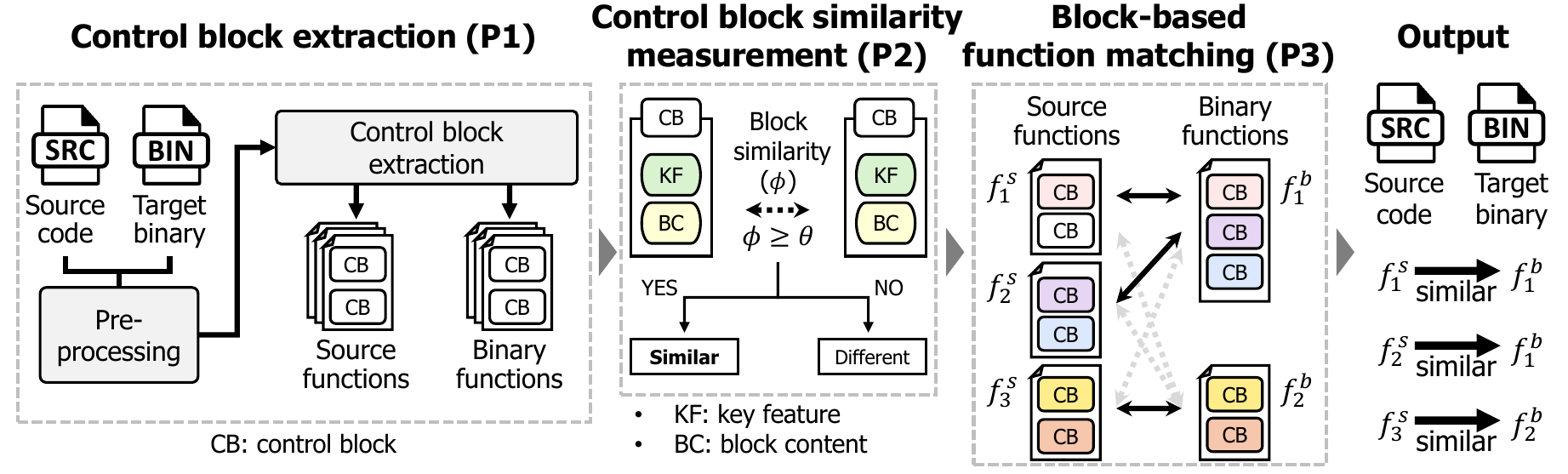}
    \caption{Overview of \sys.}
    \label{fig:overview}

    \vspace{-0.5em}
\end{figure}

\autoref{fig:overview} depicts the workflow of \sys, which comprises three main phases: \textit{control block extraction} (\textbf{P1}), \textit{control block similarity measurement} (\textbf{P2}) and \textit{block-based function matching} (\textbf{P3}).

In \textbf{P1}, given the source code and binary, \sys first extracts functions and then constructs control blocks for each function.
To bridge the gap between source and binary representations, \sys performs normalization, reducing differences between them and transforming them into a comparable format.
In \textbf{P2}, \sys computes the similarity between control blocks.
To this end, it extracts \textit{key features} that characterize each block and \textit{block contents} as auxiliary information for matching. Using these elements, \sys performs block-level matching, identifying similar control block pairs among all control block combinations between two functions.
%
In \textbf{P3}, \sys measures function similarity based on the block-level similarity scores. It then ranks candidate binary functions according to their similarity to the input source function.



\PP{Scope and assumption}
\sys operates regardless of whether the binary is stripped. Because of function inlining, a single source function may correspond to multiple binary functions (1-to-N matching), and conversely, multiple source functions may map to a single binary function (N-to-1 matching). Instead of identifying only the most similar binary function for a given source function, \sys extracts all similar binary functions based on similarity scores, focusing on the top five. 
\subsection{Control Block Extraction (P1)}\label{subsec:p1}
Given the source code and target binary, \sys first identifies the functions and subsequently extracts control blocks from each function.

\PP{Preprocessing of source code}
Given a C/C++ source code file, 
\sys first extracts functions.
However, limiting analysis to functions alone may result in the loss of crucial context, such as macro definitions. 
%
To preserve this information
\sys preprocesses the source code using the \texttt{-E} option (\eg, \texttt{gcc} \texttt{-E} \texttt{source.c}). 
While this does not perform full compilation, it ensures that macro expansions and header file inclusions are reflected in the extracted function bodies.
\sys then uses function parsers  (\eg, Ctags~\cite{ctags}) to extract function bodies from the preprocessed source code.

\PP{Preprocessing of binary code}
\sys 
extracts functions from binaries by leveraging a binary reverse engineering tool (\eg, IDA Pro~\cite{idapro}, Ghidra~\cite{ghidra}, Binary Ninja~\cite{binaryninja} and rev.ng~\cite{revng}),
which can
generate C-like low-level pseudocode for all functions within a binary. We use C-like low-level pseudocode instead of raw assembly code to more clearly identify control structures, such as loops and conditionals. This makes it easier to extract control blocks using C/C++ parsers.



\PP{Control block extraction}
\sys slices a function into finer-grained units called a \textit{control block}, 
a minimal semantic unit within a function that encapsulates functional features related to program flow control.
The underlying concept is that even when inter-function or intra-function control flow changes, the individual small units remain fundamentally stable. 
To implement this approach, \sys considers two branching patterns for control block selection: (1) \textit{internal} and (2) \textit{external}.
%


\vspace{0.5em}
\begin{itemize}
    \setlength\itemsep{0.3em}
    \item \textbf{Internal branching.} This includes conditions, else-conditions, and loop code statements. 
    Although their order and syntax may vary during compilation and decompilation, individual condition units generally preserve their syntactic structure (see \autoref{subsec:moti}).
    %

    \item \textbf{External branching.} This encompasses branching information across functions. It focuses on function call-related constructs, which we further subdivided into three distinct categories: user-defined function calls, C library invocations, and recursive function calls.
\end{itemize}
\vspace{0.5em}


Along with branching blocks, \sys also considers string literals that are resilient to compilation changes. Based on this classification, \sys categorizes control blocks into seven types.

\vspace{0.5em}
\begin{enumerate}
    \setlength\itemsep{0.3em}
    \item[\textbf{C1.}] \textbf{\texttt{Condition}}: Includes conditional expressions (\eg, \texttt{if} and \texttt{switch} conditions).

    \item[\textbf{C2.}]\textbf{\texttt{Else-Condition}}: Includes alternative branches (\texttt{else} conditions).

    \item[\textbf{C3.}]\textbf{\texttt{Loop}}: Includes loop structures (\eg, \texttt{for} and \texttt{while}).

    \item[\textbf{C4.}]\textbf{\texttt{String}}:
    Includes string literals, which mostly remain invariant across compilations.

    \item[\textbf{C5.}]\textbf{\texttt{LibcFunction}}: Includes standardized C library function calls.

    \item[\textbf{C6.}]\textbf{\texttt{CalleeFunction}}: Includes user-defined function calls.

    \item[\textbf{C7.}]\textbf{\texttt{RecurFunction}}: Includes recursive function calls.
\end{enumerate}
\vspace{0.5em}

%

\begin{table}[t]
\renewcommand{\tabcolsep}{4.9mm}
\caption{\label{table:controlblock}Types and representative components of defined control blocks.}

\small
\begin{center}
\begin{tabular}{|c|l|l|l|}
\hline
\rule{0in}{2.2ex}\textbf{IDX}
&\multicolumn{1}{c|}{\textbf{Block type}}
&\multicolumn{1}{c|}{\textbf{Key feature}}
&\multicolumn{1}{c|}{\textbf{Block content}}\\\hline\hline
\rule{0in}{2.2ex}C1
&\texttt{Condition}
&Condition expression
&Codes in conditional state\\
C2
&\texttt{Else-Condition}
&\textcolor{lightgray}{\texttt{None}}
&Codes in conditional state\\
C3
&\texttt{Loop}
&Condition expression
&Codes in loop state\\
C4
&\texttt{String}
&String literal
&\textcolor{lightgray}{\texttt{None}}\\
C5
& \texttt{LibcFunction}
& (Func. name, \#Params)
& Parameter information \\
C6
&\texttt{CalleeFunction}
&(Func. name, \#Params)
&Parameter information\\
C7
&\texttt{RecurFunction}
&(Func. name, \#Params)
&Parameter information\\\hline
\end{tabular}
\end{center}

\end{table}

Each block consists of \textit{key features} and \textit{block contents}.

\vspace{0.3em}
\begin{itemize}
    \setlength\itemsep{0.2em}
    \item \textbf{Key feature}: Elements that include the primary feature for comparing blocks.
    \item \textbf{Block content}: Elements that include an auxiliary feature for block matching.
\end{itemize}
\vspace{0.3em}


\autoref{table:controlblock} summarizes the key features and block contents of each block type.
Control blocks are selected based on certain control structures that preserve their logical intent despite syntactic transformations during compilation.
\texttt{Condition}, \texttt{Else-Condition}, and \texttt{loop} blocks (C1 to C3) 
generally maintain the structure of original condition expressions
despite potential variations in code syntax due to compilation.
The key feature of these blocks is the condition expression, while all code lines contained within each conditional block are stored as the block contents.
\sys extracts control blocks hierarchically to handle nested structures. In a nested conditional statement, the outer block has its own condition as the key feature, which may be empty if it only contains another control block. The inner block is extracted separately with its own key feature and block content.

%
\texttt{String} blocks (C4) contain string literals. Although string literals do not strictly belong to control flow, we include them as an auxiliary type of control block because string literals tend to remain relatively unchanged across compilations.

\begin{table}[b]

\centering
\caption{C library function group (referring to \cite{gnuclibrary, llvmdoxygen}).}
\label{tab:libc_groups}

\small
\begin{tabular}{|c|l|}
  \hline
  \rule{0in}{2.2ex}\textbf{Representative function} & \multicolumn{1}{c|}{\textbf{Group}} \\\hline\hline
  \rule{0in}{2.2ex}\texttt{printf} & \begin{tabular}[l]{@{}l@{}}\texttt{printf}, \texttt{iprintf}, \texttt{puts}, \texttt{putchar}\end{tabular} \\\hline
  \texttt{fprintf} & \begin{tabular}[l]{@{}l@{}}\texttt{fprintf}, \texttt{fputc}, \texttt{fputs}, \texttt{fwrite}, \texttt{fiprintf}, \texttt{\_\_small\_fprintf}\end{tabular} \\\hline
  \texttt{strncat} & \begin{tabular}[l]{@{}l@{}}\texttt{strncat}, \texttt{strcat}\end{tabular} \\\hline
  \texttt{strrchr} & \begin{tabular}[l]{@{}l@{}}\texttt{strrchr}, \texttt{strchr}, \texttt{strlen}, \texttt{strpbrk}\end{tabular} \\\hline
  \texttt{strncmp} & \begin{tabular}[l]{@{}l@{}}\texttt{strcmp}, \texttt{memcmp}, \texttt{strncmp}, \texttt{strstr}\end{tabular} \\\hline
  \texttt{memcopy} & \begin{tabular}[l]{@{}l@{}}\texttt{memcpy}, \texttt{bcopy}, \texttt{memset}, \texttt{strncpy}, \texttt{stpncpy}, \texttt{bcmp}, \texttt{memccpy}\end{tabular} \\\hline
  \texttt{gettext} & \begin{tabular}[l]{@{}l@{}}\texttt{gettext}, \texttt{dgettext}, \texttt{dcgettext}, \texttt{ngettext}, \texttt{dngettext}, \texttt{dcngettext}\end{tabular} \\\hline
  \texttt{errno} & \begin{tabular}[l]{@{}l@{}}\texttt{errno}, \texttt{\_\_errno\_location}\end{tabular} \\\hline
\end{tabular}
\end{table}

Next, although the \texttt{LibcFunction} (C5) and \texttt{CalleeFunction} blocks (C6) can be simplified during optimization, they still encapsulate critical function call information.
To extract \texttt{LibcFunction} blocks, we selected 346 standard C library functions (\eg, \texttt{printf}, \texttt{scanf}) based on the IBM Standard C Library Functions Table document, with additional inclusion of GNU gettext (i18n/l10n) and POSIX/BSD socket functions~\cite{gnuclibrary,ibmlibc}. 
However, C library functions can be optimized or replaced by alternative functions due to preprocessing options and compiler optimization. 
Specifically, the LLVM compiler often performs optimizations that replace certain C library functions with simpler equivalents when specific conditions are met, as implemented in the LLVM Library Calls Simplifier (\texttt{SimplifyLibCalls.cpp}).
For example, when \texttt{printf} is used without format specifiers and the format string ends with a newline character, it can be replaced with \texttt{puts} (\eg, \texttt{printf(``foo\textbackslash n'')} $\rightarrow$ \texttt{puts(``foo'')}). Because the optimization level of binary code is indeterminate, it is impossible to predict the level of optimization applied. 
Therefore, by consulting the relevant documentation, we grouped together functions that can be substituted, such as those described in LLVM \texttt{SimplifyLibCalls} and the GNU C Library manual~\cite{gnuclibrary, llvmdoxygen}. \autoref{tab:libc_groups} shows the representative C library function groups.

Finally, \texttt{RecurFunction} blocks (C7) are not affected by function inlining and are also considered by \sys as a type of control block.



\begin{lstlisting}[float, language=C, 
caption = {\label{lst:example}Example code for extracting control blocks.}, style=base, basicstyle=\fontsize{7.5}{8.5}\ttfamily\color{black}, deletekeywords={for}, mathescape, breakatwhitespace=true, breaklines=true]
$\text{\quad\textcolor{gray}{// KF: Key feature, BC: Block content}}$
$\text{\quad\textcolor{black}{\#include <stdlib.h>}}$

$\text{\quad\textcolor{black}{int add\_one(int x) \{ return x + 1; \}}}$
$\text{\quad\textcolor{black}{int factorial(int n);}}$
$\text{\quad\textcolor{black}{int main(void) \{}}$
$\text{\quad\;\;\;\textcolor{black}{int i = 0, p = 2, q = 0, r = 3;}}$
$\text{\quad\;\;\;\textcolor{black}{if (\colorbox{fseorange}{p <= 1}) \colorbox{fsepurple}{return 0;}\hspace{3.6em} \textcolor{gray}{//} [C1] Condition (\colorbox{fseorange}{KF}, \colorbox{fsepurple}{BC})}}$
$\text{\quad\;\;\;\textcolor{black}{else \{ \colorbox{fsepurple}{r = r + 5;} \} \hspace{4.35em} \textcolor{gray}{//} [C2] Else-Condition (\colorbox{fsepurple}{BC})}}$
$\text{\quad\;\;\;\textcolor{black}{while (\colorbox{fseorange}{i < r}) \{\hspace{6.83em} \textcolor{gray}{//} [C3] Loop (\colorbox{fseorange}{KF})}}$
$\text{\quad\;\;\;\;\;\;\textcolor{black}{\colorbox{fsepurple}{q = q + 1;} \hspace{8em} \textcolor{gray}{/*} [C3] Loop (\colorbox{fsepurple}{BC})}}$
$\text{\quad\;\;\;\;\;\;\textcolor{black}{\colorbox{fsepurple}{i++;} \hspace{12.5em} [C3] Loop (\colorbox{fsepurple}{BC}) \textcolor{gray}{*/} }}$
$\text{\quad\;\;\;\textcolor{black}{\}}}$
$\text{\quad\;\;\;\textcolor{black}{const char *s = \colorbox{fseorange}{"hi"}; \hspace{3.33em} \textcolor{gray}{//} [C4] String (\colorbox{fseorange}{KF})}}$
$\text{\quad\;\;\;\textcolor{black}{void *buf = \colorbox{fseorange}{malloc}(\colorbox{fsepurple}{16});  \hspace{2.05em} \textcolor{gray}{//} [C5] LibcFunction I (\colorbox{fseorange}{KF}, \colorbox{fsepurple}{BC})}}$
$\text{\quad\;\;\;\textcolor{black}{\colorbox{fseorange}{free}(\colorbox{fsepurple}{buf});  \hspace{8.56em} \textcolor{gray}{//} [C5] LibcFunction II (\colorbox{fseorange}{KF}, \colorbox{fsepurple}{BC})}}$
$\text{\quad\;\;\;\textcolor{black}{int t = \colorbox{fseorange}{add\_one}(\colorbox{fsepurple}{q}); \hspace{4.05em} \textcolor{gray}{//} [C6] CalleeFunction I (\colorbox{fseorange}{KF}, \colorbox{fsepurple}{BC})}}$
$\text{\quad\;\;\;\textcolor{black}{int u = \colorbox{fseorange}{factorial}(\colorbox{fsepurple}{3}); \hspace{3.07em} \textcolor{gray}{//} [C6] CalleeFunction II (\colorbox{fseorange}{KF}, \colorbox{fsepurple}{BC})}}$
$\text{\quad\;\;\;\textcolor{black}{return t + q + u + (int)s[0];}}$
$\text{\quad\textcolor{black}{\}}}$

$\text{\quad\textcolor{black}{int factorial(int n) \{}}$
$\text{\quad\;\;\;\textcolor{black}{if (\colorbox{fseorange}{n <= 0}) \colorbox{fsepurple}{return 0;}\hspace{3.6em} \textcolor{gray}{//} [C1] Condition (\colorbox{fseorange}{KF}, \colorbox{fsepurple}{BC})}}$
$\text{\quad\;\;\;\textcolor{black}{return \colorbox{fseorange}{factorial}(\colorbox{fsepurple}{n - 1}); \hspace{1.6em} \textcolor{gray}{//} [C7] RecurFunction (\colorbox{fseorange}{KF}, \colorbox{fsepurple}{BC})}}$
$\text{\quad\textcolor{black}{\}}}$
\end{lstlisting}

When we examined the proportion of lines belonging to control blocks across all collected source and binary functions in our experimental setup (see \autoref{subsec:rq1}), we observed that more than 82.66\% of source lines and over 73.77\% of binary lines could be mapped to block types.
Code lines that do not fall into these categories are mostly simple variable assignments, and considering that variable names are not particularly helpful for cross-comparison, 
the proportion of control blocks
can be regarded as sufficiently high.
The slightly lower coverage in binaries is primarily due to the decompilation process, which introduces a large number of trivial variable assignments. However, these assignments do not contribute meaningful information when mapping to the source code, and thus considering them would even degrade performance. This observation highlights that control blocks provide a more reasonable and effective abstraction for source-to-binary matching.



For example, \autoref{lst:example} illustrates the process of block extraction.
As in the case of a \texttt{Condition} block (\ie, line \#8), the block contents may consist of a single line of code, whereas, as in the case of a \texttt{Loop} block (\ie, lines \#10 to \#12), it may span multiple lines.
If a function makes a recursive call, it is defined as a \texttt{RecurFunction} (\ie, line \#24), while calls to other functions are defined as \texttt{CalleeFunction} (\ie, lines \#17 and \#18). A single line of code can also belong to multiple control blocks. For example, if there is a statement ``\texttt{printf(``hello'');}'', the string literal \texttt{``hello''} belongs both to a \texttt{String} block and to the block content of a \texttt{LibcFunction}.

\PP{Function normalization}
\sys further applies function normalization, ensuring more precise function matching.
To ensure the generality of \sys, we avoid applying overly heuristic normalization rules and instead focus on clear transformations that address changes frequently introduced during compilation. \sys first (1) removes all comments from the source functions. Next, it (2) replaces all \texttt{ASCII} and hexadecimal characters with their decimal representations, to account for the frequent character-type conversions that occur during compilation (\eg, \texttt{0x5E} $\rightarrow$ \texttt{94}). 
Finally, (3) every occurrence of variable names and parameter names that are not used in cross-comparison is replaced with the keywords \texttt{LVAR} and \texttt{PARAM}, respectively.
For all blocks, both the key features and block content are preserved in the normalized form.

\begin{table}[t]
\renewcommand{\tabcolsep}{0.9mm}
\caption{\label{table:feature_lst}List of features used for internal branching block similarity measurement.}
\vspace{-1em}
\small
\begin{center}
\begin{tabular}{|c|c|c|ccc|ccc|c|cccll|}
\hline
\rule{0in}{2.2ex}\textbf{Index}   
& 0                                       
& 1                     
& \multicolumn{1}{c|}{$\;\;$2$\;\;$}
& \multicolumn{1}{c|}{$\;$...$\;$}        
& $\;\;$16$\;\;$       
& \multicolumn{1}{c|}{$\;\;$17$\;\;$}           
& \multicolumn{1}{c|}{$\;$...$\;$}           
& $\;\;$362$\;\;$           
& 363           
& \multicolumn{1}{c|}{$\;\;$364$\;\;$}     
& \multicolumn{1}{c|}{$\;$...$\;$}     
& \multicolumn{3}{c|}{n}\\\hline\hline
\rule{0in}{2.2ex}\textbf{Info.} 
& \begin{tabular}[c]{@{}c@{}}Local\\variable used\end{tabular} 
& \begin{tabular}[c]{@{}c@{}}Parameter\\variable used\end{tabular}
& \multicolumn{3}{c|}{\begin{tabular}[c]{@{}c@{}}Operator\\used\end{tabular}} 
& \multicolumn{3}{c|}{\begin{tabular}[c]{@{}c@{}}C library\\function used\end{tabular}} 
& \begin{tabular}[c]{@{}c@{}}Unique UDF*\\ count match\end{tabular} 
& \multicolumn{5}{c|}{\begin{tabular}[c]{@{}c@{}}Unique string/\\ number/functions\end{tabular}} \\\hline
\end{tabular}
\end{center}

{\raggedleft\scriptsize UDF*: User-Defined Function\par}
\vspace{-0.4em}
\end{table}

\subsection{Control Block Similarity Measurement (P2)}\label{subsec:p2}
Next, we introduce \sys's approach for computing similarity between two blocks.
Here, comparisons between block types that have little chance of mapping (\eg, \texttt{Condition} \textit{vs.} \texttt{String}) can impair the performance and accuracy.
%
Therefore,
\sys compares blocks of the similar type: it compares C1, C2, and C3 with each other (called \textit{internal branching} blocks), compares C5, C6, and C7 (called \textit{external branching} blocks), and compares C4 (\textit{string} blocks) separately.
%
%

\sys applies fine-grained similarity analysis to \textit{internal-branching} blocks, as they span multiple lines and embody diverse control-flow semantics with conditional variations. In contrast, \textit{external-branching} and string blocks are typically short (\ie, one or two lines), thus 
\sys adopts a categorical scheme for these blocks: exact match based on key features (similarity = 1), partial match based on block contents (similarity = 0.5), and no match (similarity = 0).

\PP{Internal branching blocks: Condition, Else-Condition, and Loop}
For internal branching blocks (C1 to C3), \sys employs a \textit{feature-based vector representation} approach that abstracts compilation-specific variations to focus on features essential for accurate matching.

\autoref{table:feature_lst} shows the overall vector structure
that consists of two components: \textit{static components} that capture predetermined features (indices 0 - 363) and \textit{dynamic components} that contain context-specific information (indices 364 - $n$).
All values in the vector are set using binary encoding (0 or 1), because the frequency of certain elements may vary during compilation and decompilation. This approach emphasizes existence rather than frequency, while also enhancing computational efficiency.

For each internal branching block, separate vectors are generated for its key features and for its block content. The details for each element are as follows.

\begin{itemize}
    \setlength\itemsep{0.3em}
    \item \textbf{Local variable used.} The value is set to 1 if a local variable is used; otherwise, it is 0.

    \item \textbf{Parameter variable used.} 
    The value is set to 1 if a parameter value is used, and 0 otherwise.

    \item \textbf{Operator used.} These elements capture operators. However, an operator can be replaced with another expression that carries a similar semantic meaning during compilation. Therefore, we group semantically similar operators (see \autoref{table:operator}), and set each vector index to 1 if an operator from the corresponding group is used, and 0 otherwise.

    \item \textbf{C library function used.} 
    For the 346 predefined standard C library functions (see \autoref{subsec:p1}), each corresponding vector entry is set to 1 if the function is used and 0 otherwise.

    \item \textbf{Unique user-defined function count match.} Whether the count of unique user-defined functions matches between the compared blocks.
    This feature is computable only by comparing a source block with a candidate binary block. For internal-branching blocks extracted from the source function, \sys fixes the value to 1. For internal-branching blocks in the candidate binary function, \sys sets the value to 1 if the number of distinct user-defined functions in the binary block equals that of the source block; otherwise, 0.

    \item \textbf{Unique string/number/functions.} Three important but non-fixed elements (\ie, strings, numeric information, and invoked function information) are additionally incorporated into the vector dynamically. 
    For each of these elements, a new vector entry is created and its value is set to 1 if present; otherwise, it remains 0 when absent from the current vector but present in the compared one.
    In practice, only a small number of these elements are added, with only a marginal increase in vector size and negligible overhead during block similarity computation.
    
\end{itemize}

\begin{table}[t]
\centering
  \caption{\label{table:operator}Operator groups in condition expressions (14 groups).}
  
\hspace{-2.7em}\begin{minipage}{0.42\linewidth}
  \small
  \centering
  \vspace{-0.7em}
  \begin{tabular}{|c|l|}
  \hline  
  \rule{0in}{2.2ex}\textbf{Group}&\multicolumn{1}{c|}{\textbf{Operators}}\\\hline\hline
  \rule{0in}{2.2ex}\texttt{equOpr}&\texttt{equals}, \texttt{notEquals}\\\hline
 \texttt{comOpr}&\begin{tabular}[l]{@{}l@{}}\texttt{lessThan}, \texttt{greaterThan},\\
\texttt{lessEqualsThan},\\ \texttt{greaterEqualsThan}\end{tabular}
\\\hline
\texttt{addOpr}&\begin{tabular}[l]{@{}l@{}}\texttt{addition}, \texttt{assignmentPlus}\end{tabular}
\\\hline
\texttt{subOpr}&\begin{tabular}[l]{@{}l@{}}\texttt{subtraction}, \texttt{assignmentMinus}\end{tabular}
\\\hline
\texttt{mulOpr}&\begin{tabular}[l]{@{}l@{}}\texttt{multiplication},\\ \texttt{assignmentMultiplication}\end{tabular}
\\\hline
\texttt{divOpr}&\begin{tabular}[l]{@{}l@{}}\texttt{division}, \texttt{assignmentDivision}\end{tabular}
\\\hline
\texttt{modOpr}&\begin{tabular}[l]{@{}l@{}}\texttt{modulo}, \texttt{assignmentModulo}\end{tabular}
\\\hline
\texttt{notOpr}&\begin{tabular}[l]{@{}l@{}}\texttt{not}, \texttt{logicalNot}\end{tabular}
\\\hline
\end{tabular}
\end{minipage}\hspace{1.2em}
\begin{minipage}{0.46\linewidth}
  \small
  \centering
 
  \vspace{-0.7em}
  \begin{tabular}{|c|l|}
  \hline  
  \rule{0in}{2.2ex}\textbf{Group}&\multicolumn{1}{c|}{\textbf{Operators}}\\\hline\hline
 \rule{0in}{2.2ex}\texttt{xorOpr}&\begin{tabular}[l]{@{}l@{}}\texttt{xor}, \texttt{logicalXor}\end{tabular}
\\\hline
\texttt{shiftOpr}&\begin{tabular}[l]{@{}l@{}}\texttt{shiftLeft}, 
\texttt{arithmeticShiftLeft},\\
\texttt{shiftRight}, \texttt{arithmeticShiftRight}, \\
\texttt{assignmentArithmeticShiftLeft}, \\\texttt{assignmentArithmeticShiftRight}
\end{tabular}
\\\hline
%
\texttt{logAndOrOpr}&\begin{tabular}[l]{@{}l@{}}\texttt{logicalAnd}, \texttt{logicalOr}\end{tabular}
\\\hline
\texttt{bitAndOrOpr}&\begin{tabular}[l]{@{}l@{}}\texttt{or}, \texttt{and}\end{tabular}
\\\hline
\texttt{accOpr}&\begin{tabular}[l]{@{}l@{}}\texttt{indirectFieldAccess}, \texttt{fieldAccess}, \\\texttt{indirectIndexAccess}\end{tabular}
\\\hline
\texttt{infOpr}&\begin{tabular}[l]{@{}l@{}}\texttt{infiniteLoop}\end{tabular}
\\\hline
\texttt{assignOpr}&\begin{tabular}[l]{@{}l@{}}\texttt{assignment}\end{tabular}
\\\hline
%

\end{tabular}
\end{minipage}

\vspace{-0.5em}
\end{table}

To compute the similarity between two blocks, \sys compares the vectors extracted from their key features and block contents, respectively. Given two vectors to be compared, \sys first performs feature space alignment: If one vector contains dynamically added entries that are absent in the other, the same entries are appended to the latter with their values set to 0.

\sys then measures the similarity between the two vectors, denoted as $\mathbf{v}_X$ and $\mathbf{v}_Y$, using \textit{cosine similarity} ($\phi$). The cosine similarity between two vectors is calculated as follows.

\vspace{0.6em} \begin{center} \small $\phi(\mathbf{v}_X, \mathbf{v}_Y) = \displaystyle\frac{\mathbf{v}_X \cdot \mathbf{v}_Y}{||\mathbf{v}_X|| \,||\mathbf{v}_Y||} = \displaystyle\frac{\sum_{i=1}^m (\mathbf{v}_X)_i\times (\mathbf{v}_Y)_i}{\sqrt{\sum_{i=1}^m ((\mathbf{v}_X)_i)^2} \times \sqrt{\sum_{i=1}^m ((\mathbf{v}_Y)_i)^2}}$ \end{center} \vspace{0.7em}

Using this approach, \sys computes the similarity of the key feature vectors ($\phi_k$) and the similarity of the block content vectors ($\phi_b$) for each pair of internal branching blocks. Let the two blocks to be compared be denoted as 
$B_{I_1}$ and $B_{I_2}$.
The final block similarity ($\Phi_{b}$) is defined as the average of the key feature similarity and the block content similarity. 
\vspace{0.6em}
\[
\Phi_{b}(B_{I_1}, B_{I_2}) = (\phi_k + \phi_b)/2
\]

Although key features typically consist of a single line, block contents may span multiple lines. Averaging both measures emphasizes the importance of key features, which capture the core semantics despite their brevity.
For \texttt{Else-Condition} blocks, which do not include a key
feature, only the similarity derived from the block content is used.

\begin{figure}[t]
    \centering
    \includegraphics[width=1\textwidth]{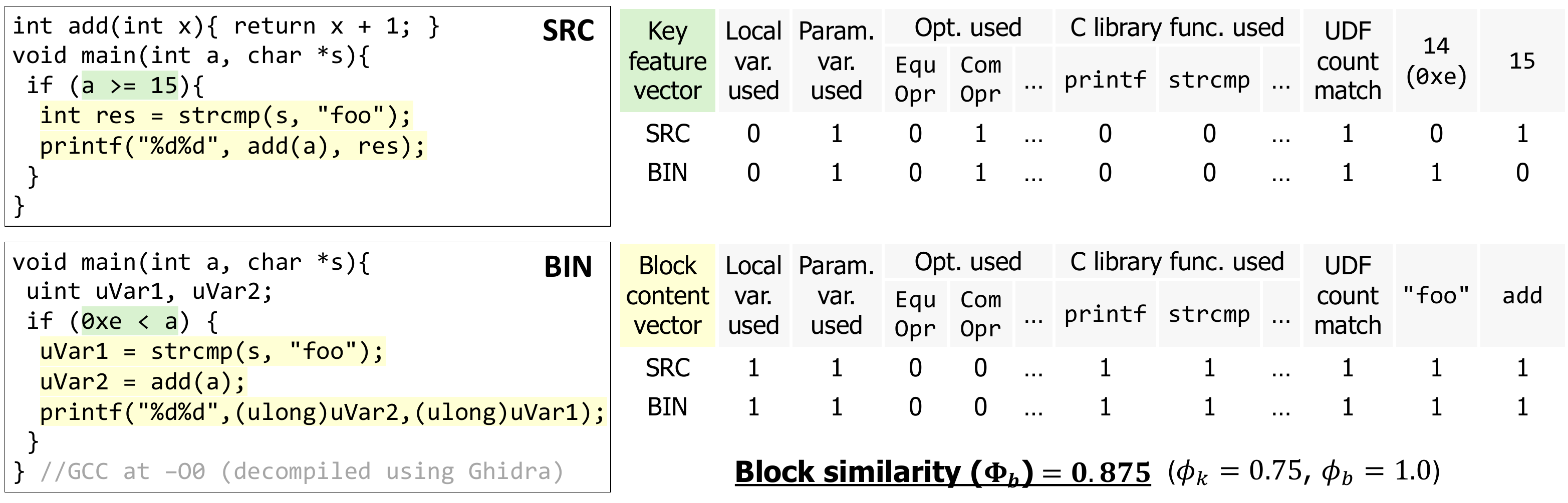}
    \vspace{-0.5em}
    \caption{Flow of internal branching block vector extraction and similarity comparison for the example code.} 
    \label{fig:p2_ex}

    \vspace{-1em}
\end{figure}

\autoref{fig:p2_ex} illustrates \sys's similarity computation on the example code. Although the conditional expressions differ (\texttt{a} \texttt{>=} \texttt{15} and \texttt{0xe} \texttt{<} \texttt{a}) and most block statements are modified, \sys's approach produced a similarity of 0.875. In contrast, 
cosine similarity computed on space-separated tokens and \texttt{Levenshtein} distance–based similarity yielded similarity scores of 0.3640 and 0.4863, respectively,  highlighting \sys's effectiveness in cross-domain block similarity analysis.

\PP{External branching blocks: LibcFunction, CalleeFunction, and RecurFunction}
Here, \sys first examines the key features by comparing the function name and the number of parameters. If both match, the block similarity $\Phi_{b} = 1$.
However, when the binary is stripped, function names may be altered. 
Therefore, if the key features do not match, 
%
\sys examines the parameter information (\ie, block contents). 
Considering the order, if both the parameter values and types match, we set $\Phi_{b} = 1$; if only the types match, $\Phi_{b} = 0.5$; otherwise, $\Phi_{b} = 0$.

\PP{String blocks.}
For \texttt{string} blocks (C4), because string literals rarely change, they are directly compared. If they match, the block similarity $\Phi_{b} = 1$; otherwise, $\Phi_{b} = 0$.
%


\subsection{Control Block-Based Function Matching (P3)}\label{subsec:p3}
\sys computes initial function similarity ($\Phi_{f}$) by counting the number of similar blocks between the source function ($f_s$) and the binary function ($f_b$). Let $\theta_b$ be the threshold for block similarity.
For function similarity computation, each source block is matched to at most one binary block. If multiple binary blocks exceed $\theta_b$ for a source block, \sys selects the highest-similarity pair, and the matched source and binary blocks are excluded from subsequent pairing to avoid duplicate matches.
\vspace{0.5em}
\[\Phi_{f}(f_s, f_b) = \dfrac{|\{\,B_s \in f_s \mid \exists\, B_b \in f_b,\;\Phi_{b}(B_s, B_b) \geq \theta_b|}{|\{\,B_s \in f_s\}|}
\]


\PP{Addressing function inlining}
SBridge handles function inlining through \textit{length-based weighting}. Although large length differences between functions usually lower similarity, \sys avoids penalizing cases where inlining makes a binary function much larger than its source, since the source is still fully included.


To this end, \sys employs a structured matching strategy based on the source call graph. Starting from the root, it performs top-down matching via depth-first search (DFS), aligning binary functions with their source counterparts. 
When a match is found, two conditions are checked: (1) whether the source function has child nodes, and (2) whether the binary function includes external calls (C5 block). If both hold, this indicates a low probability of inlining; if the source has children but the binary has no external calls, this indicates a high probability of inlining, suggesting that the child may have been merged into the parent.

To support this process, \sys defines the following function-length weighting mechanism ($w_{length}$). In this equation, \texttt{\#blocks}($f$) represents the number of control blocks in the function $f$.

\vspace{0.5em}
\begin{center}
    $w_{length} = \dfrac{1.0}{1.0 + \log \big(\text{\texttt{\#blocks}}(f_b)/\text{\texttt{\#blocks}}(f_s) + 1.0 \big)}$
\end{center}
\vspace{0.5em}

%
%

A logarithmic function is applied to mitigate extreme differences in the number of blocks between binary and source functions, where the ``+1.0'' term ensures valid computation even when \texttt{\#blocks}($f_b$) is zero. If only one source function exists or no call relationships are present, \sys sets $w_{length}=1$. When the preceding analysis indicates a low probability of inlining and the binary function has substantially more blocks, $w_{length}$ decreases to reduce the similarity score. In contrast, when the analysis suggests inlining, \sys fixes $w_{length}=1$ so that length differences do not affect the computation. \autoref{sec:dis} provides a more detailed discussion on function inlining.

\PP{Function similarity} 
The final function similarity $\Phi$ 
is calculated as follows.

\vspace{0.5em}
\begin{center}
    $\Phi(f_s, f_b) = w_{length} \cdot \Phi_{f}$
\end{center}
\vspace{0.5em}

Based on these similarity scores, \sys ranks candidate binary functions and identifies the binary function that is most similar to the input source function.

\section{Evaluation}

In this section, we evaluate \sys based on the following four questions.

\begin{itemize}
    \setlength\itemsep{0.2em}
    \item [\textbf{RQ1.}] \textbf{Accuracy:} 
    How accurately does \sys detect binary counterparts of source functions?
    \item [\textbf{RQ2.}] \textbf{Efficacy:} How effectively does \sys handle function inlining?
    \item [\textbf{RQ3.}] \textbf{Performance:} 
    How fast and scalable is \sys in detecting similar functions?
    
    \item [\textbf{RQ4.}] \textbf{Application:}
    How effective is \sys when used for vulnerability detection? 
\end{itemize}

\sys is implemented in F\# with 4.1K lines of code (LOC). It uses Ctags~\cite{ctags} and Joern parser~\cite{yamaguchi2014modeling} for function extraction and normalization.
Although compatible with various binary analysis tools,
\sys adopts Ghidra~\cite{ghidra}, an open-source tool, to avoid licensing constraints.
Experiments ran on
Ubuntu 22.04, Intel Core Ultra 7 265K processor clocked at 3.9GHz, 32GB RAM, and 1TB SSD.

%





\subsection{Accuracy of \sys}\label{subsec:rq1}
\PP{Dataset}
We used the binary code similarity analysis (BCSA) benchmark from BinKit~\cite{binkit}.
We selected the top two GNU software packages in BinKit with the most binaries: GNU Coreutils (v9.0) and Inetutils (v2.4).
Initially, 122 target packages were collected from this benchmark. 

To evaluate \sys across diverse environments, we compiled each binary under 32 configurations, derived from four architectures (\texttt{ARM32}, \texttt{ARM64}, \texttt{x86}, \texttt{x64}), two compilers (\texttt{GCC} 9.4.0 and \texttt{Clang} 13.0), two optimization levels (\texttt{-O0} and \texttt{-O2}), and two symbol management options (\texttt{strip} and \texttt{no\_strip}). This produced 3,904 binaries containing 624,192 functions for accuracy evaluation.

\PP{Methodology}
We evaluate \sys by using the source codes of 122 collected packages, comprising 1,618 source functions, as input. The goal is to assess how accurately \sys identifies source functions ($f_s$) that have been compiled into binaries ($f_b$) under various environments. 
If \sys identifies $f_b$ as similar to $f_s$, it is considered a true positive (TP). 
%
For ground truth, we extracted function mappings from \texttt{no\_strip} and \texttt{O0} binaries based on function names. When multiple functions shared the same name, we manually examined the source and binary functions. 
In this process,
we assume functions in \texttt{O0} binaries are not inlined. 
Although this assumption may introduce minor discrepancies, it ensures a consistent basis for comparison across all tools.
%
%
In our experiments, we set the block similarity threshold $\theta_b$ to 0.7 (see \autoref{subsec:p3}), and analyze its sensitivity at the end of the accuracy evaluation.

\PP{Comparison targets}
For accuracy evaluation, we compared \sys with two state-of-the-art approaches representing the source-to-source and source-to-binary domains, both of which are publicly available and thus suitable for direct comparison:
MRT-OAST~\cite{yu2025multiple} (source-to-source) and BinaryAI~\cite{jiang2024binaryai} (source-to-binary).
MRT-OAST is a deep learning–based code clone detection technique that leverages optimized abstract syntax trees (OAST) and a Siamese network.
We employed MRT-OAST to identify code clones between input source code and decompiled binary code, and then measured its accuracy.
BinaryAI~\cite{jiang2024binaryai} detects reused components by mapping binary functions to source functions. For our evaluation, we adapt BinaryAI for function-level evaluation by reversing its process. Specifically, when it identifies a source function ($f_s$) similar to a binary function ($f_b$), we invert the mapping ($f_s \rightarrow f_b$) and use this to compare against \sys. The potential limitations of this adaptation are discussed in \autoref{sec:dis}.

Both approaches can be applied to detecting modified code clones, thereby addressing our first challenge (\ie, loss of source code context) to some extent. However, MRT-OAST falls short in tackling the second and third challenges (\ie, architecture and compiler variability and addressing function inlining), while BinaryAI does not sufficiently consider inlining and thus struggles with the third challenge. For these reasons, we considered them appropriate baselines for comparison.

\vspace{-0.1em}
\PP{Evaluation metric}
We use \texttt{Recall@1},  \texttt{Recall@5}, and Mean Reciprocal Rank (\texttt{MRR}) for accuracy evaluation (TP = true positive, FN = false negative). 

\vspace{0.5em}
\begin{center}
    $\text{\texttt{Recall@1}} = \dfrac{\text{\texttt{\#TP@1}}}{\text{\texttt{\#TP@1}} + \text{\texttt{\#FN@1}}}$,\hspace{1em} $\text{\texttt{Recall@5}} = \dfrac{\text{\texttt{\#TP@5}}}{\text{\texttt{\#TP@5}} + \text{\texttt{\#FN@5}}}$,\hspace{1em}
    $\text{\texttt{MRR}} = \dfrac{1}{|Q|} \displaystyle\sum_{i=1}^{|Q|} \dfrac{1}{rank_i}
    $
\end{center}
\vspace{0.5em}

\texttt{TP@1} denotes that the correct binary function (\ie, compiled from the source function) is ranked first for the given source function, while \texttt{TP@5} means it appears within the top five. Otherwise, the case is regarded as \texttt{FN@1} or \texttt{FN@5}, respectively.
%
Because each source function is evaluated against only the top-$k$ candidates, an incorrect match (false positive; FP) necessarily means the correct one is missed (FN). Hence, FP and FN coincide, and \texttt{Recall@k} is used as the evaluation metric~\cite{jiang2024binaryai}.
\texttt{MRR} is used to evaluate how highly the correct results are ranked for a given query. In our setting, the query is an input source function, and the correct result is its corresponding binary function.

\begin{table}[t]
\renewcommand{\tabcolsep}{2.3mm}
\caption{\label{table:acc}The accuracy measurement results 
by architecture, compiler, optimization, and symbol management
for the three tools. 
We consider only input source functions for which the ground truth could be measured. 
The bold results indicate the highest recall for each configuration
(\texttt{R@1}: \texttt{Recall@1}, \texttt{R@5}: \texttt{Recall@5}).}
\vspace{-1em}
\small
\begin{center}

\begin{tabular}{cccccccccc}
\hline
\multicolumn{1}{|c||}{\multirow{2}{*}{\rule{0in}{2.2ex}}} 
& \multicolumn{3}{c||}{\textbf{MRT-OAST}}                       
& \multicolumn{3}{c||}{\textbf{BinaryAI}}                       
& \multicolumn{3}{c|}{\textbf{\sys}}\\\cline{2-10}    
\multicolumn{1}{|c||}{\rule{0in}{2.2ex}}
& \multicolumn{1}{c|}{\texttt{\textbf{R@1}}} 
& \multicolumn{1}{c|}{\texttt{\textbf{R@5}}} 
& \multicolumn{1}{c||}{\texttt{\textbf{MRR}}}
& \multicolumn{1}{c|}{\texttt{\textbf{R@1}}} 
& \multicolumn{1}{c|}{\texttt{\textbf{R@5}}} 
& \multicolumn{1}{c||}{\texttt{\textbf{MRR}}}
& \multicolumn{1}{c|}{\texttt{\textbf{R@1}}} 
& \multicolumn{1}{c|}{\texttt{\textbf{R@5}}}
& \multicolumn{1}{c|}{\texttt{\textbf{MRR}}}\\\hline\hline
\multicolumn{10}{|l|}{\rule{0in}{2.2ex}\textit{By architecture}}\\\hline
\multicolumn{1}{|c||}{\rule{0in}{2.2ex}\texttt{ARM32}}
& \multicolumn{1}{r|}{0.0918}
& \multicolumn{1}{r|}{0.2188}
& \multicolumn{1}{r||}{0.1651} 
& \multicolumn{1}{r|}{0.4567}
& \multicolumn{1}{r|}{0.4994}
& \multicolumn{1}{r||}{0.4763}
& \multicolumn{1}{r|}{\textbf{0.7082}}
& \multicolumn{1}{r|}{\textbf{0.7787}}
& \multicolumn{1}{r|}{\textbf{0.7718}}\\
\multicolumn{1}{|c||}{\texttt{ARM64}}
& \multicolumn{1}{r|}{0.1270}
& \multicolumn{1}{r|}{0.2969}
& \multicolumn{1}{r||}{0.2149} 
& \multicolumn{1}{r|}{0.4364}
& \multicolumn{1}{r|}{0.5045}
& \multicolumn{1}{r||}{0.4679}
& \multicolumn{1}{r|}{\textbf{0.7751}}
& \multicolumn{1}{r|}{\textbf{0.8325}}
& \multicolumn{1}{r|}{\textbf{0.8275}}\\
\multicolumn{1}{|c||}{\texttt{x86}}
& \multicolumn{1}{r|}{0.1276}
& \multicolumn{1}{r|}{0.2885}
& \multicolumn{1}{r||}{0.2127} 
& \multicolumn{1}{r|}{0.4175}
& \multicolumn{1}{r|}{0.4979}
& \multicolumn{1}{r||}{0.4554}
& \multicolumn{1}{r|}{\textbf{0.7394}}
& \multicolumn{1}{r|}{\textbf{0.7921}}
& \multicolumn{1}{r|}{\textbf{0.7869}}\\
\multicolumn{1}{|c||}{\texttt{x64}}
& \multicolumn{1}{r|}{0.1360}
& \multicolumn{1}{r|}{0.3115}
& \multicolumn{1}{r||}{0.2280} 
& \multicolumn{1}{r|}{0.4220}
& \multicolumn{1}{r|}{0.5062}
& \multicolumn{1}{r||}{0.4618}
& \multicolumn{1}{r|}{\textbf{0.7827}}
& \multicolumn{1}{r|}{\textbf{0.8358}}
& \multicolumn{1}{r|}{\textbf{0.8299}}\\\hline\hline
\multicolumn{10}{|l|}{\rule{0in}{2.2ex}\textit{By compilers}}\\\hline
\multicolumn{1}{|c||}{\rule{0in}{2.2ex}\texttt{GCC}}
& \multicolumn{1}{r|}{0.1276}
& \multicolumn{1}{r|}{0.2906}
& \multicolumn{1}{r||}{0.2148} 
& \multicolumn{1}{r|}{0.4329}
& \multicolumn{1}{r|}{0.5099}
& \multicolumn{1}{r||}{0.4688}
& \multicolumn{1}{r|}{\textbf{0.7537}}
& \multicolumn{1}{r|}{\textbf{0.8097}}
& \multicolumn{1}{r|}{\textbf{0.8051}}\\
\multicolumn{1}{|c||}{\texttt{Clang}}
& \multicolumn{1}{r|}{0.1137}
& \multicolumn{1}{r|}{0.2672}
& \multicolumn{1}{r||}{0.1955}  
& \multicolumn{1}{r|}{0.4333}
& \multicolumn{1}{r|}{0.4941}
& \multicolumn{1}{r||}{0.4619}
& \multicolumn{1}{r|}{\textbf{0.7490}}
& \multicolumn{1}{r|}{\textbf{0.8099}}
& \multicolumn{1}{r|}{\textbf{0.8029}}\\\hline\hline
\multicolumn{10}{|l|}{\rule{0in}{2.2ex}\textit{By optimizations}}\\\hline
\multicolumn{1}{|c||}{\rule{0in}{2.2ex}\texttt{O0}}
& \multicolumn{1}{r|}{0.1672}
& \multicolumn{1}{r|}{0.3583}
& \multicolumn{1}{r||}{0.2653}  
& \multicolumn{1}{r|}{0.5087}
& \multicolumn{1}{r|}{0.5933}
& \multicolumn{1}{r||}{0.5482}
& \multicolumn{1}{r|}{\textbf{0.8496}}
& \multicolumn{1}{r|}{\textbf{0.8754}}
& \multicolumn{1}{r|}{\textbf{0.8802}}\\
\multicolumn{1}{|c||}{\texttt{O2}}
& \multicolumn{1}{r|}{0.0741}
& \multicolumn{1}{r|}{0.1995}
& \multicolumn{1}{r||}{0.1450} 
& \multicolumn{1}{r|}{0.3560}
& \multicolumn{1}{r|}{0.4088}
& \multicolumn{1}{r||}{0.3808}
& \multicolumn{1}{r|}{\textbf{0.6531}}
& \multicolumn{1}{r|}{\textbf{0.7441}}
& \multicolumn{1}{r|}{\textbf{0.7279}}\\\hline\hline
\multicolumn{10}{|l|}{\rule{0in}{2.2ex}\textit{By symbol managements}}\\\hline
\multicolumn{1}{|c||}{\rule{0in}{2.2ex}\texttt{no\_strip}}
& \multicolumn{1}{r|}{0.1310}
& \multicolumn{1}{r|}{0.2992}
& \multicolumn{1}{r||}{0.2189} 
& \multicolumn{1}{r|}{0.4378}
& \multicolumn{1}{r|}{0.5092}
& \multicolumn{1}{r||}{0.4713}
& \multicolumn{1}{r|}{\textbf{0.7909}}
& \multicolumn{1}{r|}{\textbf{0.8470}}
& \multicolumn{1}{r|}{\textbf{0.8375}}\\
\multicolumn{1}{|c||}{\texttt{strip}}
& \multicolumn{1}{r|}{0.1102}
& \multicolumn{1}{r|}{0.2586}
& \multicolumn{1}{r||}{0.1914} 
& \multicolumn{1}{r|}{0.4284}
& \multicolumn{1}{r|}{0.4948}
& \multicolumn{1}{r||}{0.4594}
& \multicolumn{1}{r|}{\textbf{0.7119}}
& \multicolumn{1}{r|}{\textbf{0.7725}}
& \multicolumn{1}{r|}{\textbf{0.7706}}\\\hline\hline
\multicolumn{10}{|l|}{\rule{0in}{2.2ex}\textit{Total result (average)}}\\\hline
\multicolumn{1}{|c||}{\rule{0in}{2.2ex}\texttt{Total}}
& \multicolumn{1}{r|}{0.1206}
& \multicolumn{1}{r|}{0.2789}
& \multicolumn{1}{r||}{0.2052}
& \multicolumn{1}{r|}{0.4331}
& \multicolumn{1}{r|}{0.5020}
& \multicolumn{1}{r||}{0.4653}
& \multicolumn{1}{r|}{\textcolor{tomato}{\textbf{0.7513}}}
& \multicolumn{1}{r|}{\textcolor{tomato}{\textbf{0.8098}}}
& \multicolumn{1}{r|}{\textcolor{tomato}{\textbf{0.8040}}}\\\hline

\end{tabular}
\end{center}
\end{table}

\begin{figure}[t!]
    \centering
    \begin{subfigure}{0.327\textwidth}
        \centering
        \includegraphics[width=\linewidth]{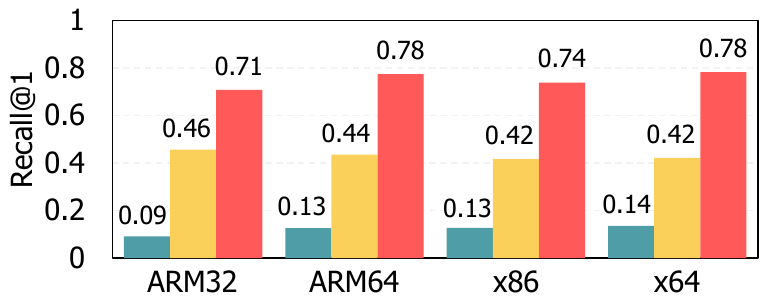}
        \caption{\label{subfig:g1}By architecture.}
    \end{subfigure}\hspace{0.2em}%
    \begin{subfigure}{0.19\textwidth}
        \centering
        \includegraphics[width=\linewidth]{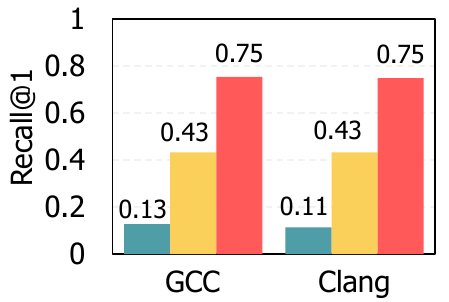}
        \caption{\label{subfig:g2}By compiler.}
    \end{subfigure}\hspace{0.2em}%
    \begin{subfigure}{0.19\textwidth}
        \centering
        \includegraphics[width=\linewidth]{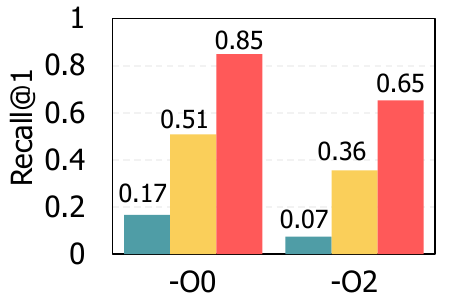}
        \caption{\label{subfig:g3}By optimization.}
    \end{subfigure}\hspace{0.2em}%
    \begin{subfigure}{0.26\textwidth}
        \centering
        \includegraphics[width=\linewidth]{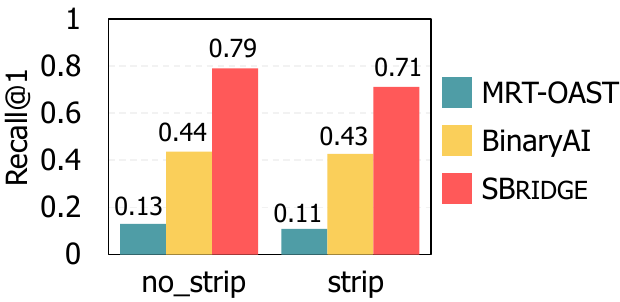}
        \caption{\label{subfig:g4}By symbol management.}
    \end{subfigure}
    \vspace{-0.5em}
    \caption{\label{fig:rec1}\texttt{Recall@1} measurement results by architecture, compiler, optimization, and symbol management.}

    \vspace{-1em}
\end{figure}
\begin{figure}[t]
    \centering
    \begin{subfigure}{0.327\textwidth}
        \centering
        \includegraphics[width=\linewidth]{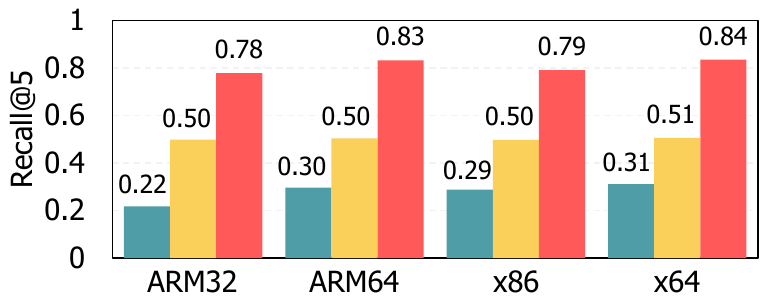}
        \caption{\label{subfig:g1}By architecture.}
    \end{subfigure}\hspace{0.2em}%
    \begin{subfigure}{0.19\textwidth}
        \centering
        \includegraphics[width=\linewidth]{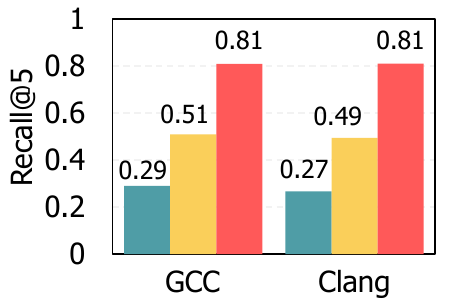}
        \caption{\label{subfig:g2}By compiler.}
    \end{subfigure}\hspace{0.2em}%
    \begin{subfigure}{0.19\textwidth}
        \centering
        \includegraphics[width=\linewidth]{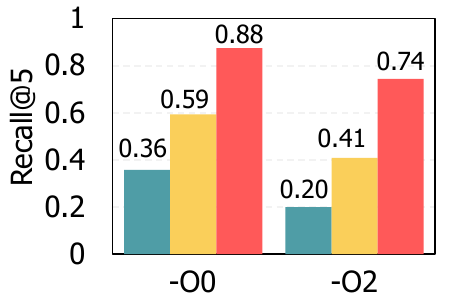}
        \caption{\label{subfig:g3}By optimization.}
    \end{subfigure}\hspace{0.2em}%
    \begin{subfigure}{0.26\textwidth}
        \centering
        \includegraphics[width=\linewidth]{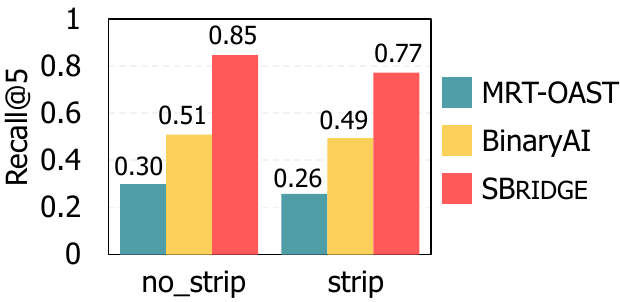}
        \caption{\label{subfig:g4}By symbol management.}
    \end{subfigure}
    \vspace{-0.5em}
    \caption{\label{fig:rec5}\texttt{Recall@5} measurement results by architecture, compiler, optimization, and symbol management.}

    \vspace{-1em}
\end{figure}

\PP{Result overview.}
Experimental results show that \sys outperforms both the MRT-OAST and BinaryAI across most configurations.
In particular, \sys achieved 6.23 times higher \texttt{Recall@1} compared to the MRT-OAST and 1.73 times higher \texttt{Recall@1} compared to BinaryAI.
\autoref{table:acc}, \autoref{fig:rec1}, and \autoref{fig:rec5} illustrate the detection results of MRT-OAST, BinaryAI, and \sys.

\vspace{-2px}
\PP{Result analysis: existing approaches}
First, we observed that function inlining has the most significant impact on the accuracy of existing approaches. This is evident from the optimization experiments: the \texttt{Recall@1} of MRT-OAST decreased from 0.1672 (at \texttt{-O0}) to 0.0741 (at \texttt{-O2}), while that of BinaryAI dropped from 0.5087 to 0.3560 (see \autoref{subfig:g3}).

In particular, MRT-OAST relies on syntactic information, and thus its identification accuracy decreases when binaries are stripped. Specifically, its \texttt{Recall@1} dropped from 0.1310 in the \texttt{no\_strip} setting to 0.1102 in the \texttt{strip} setting.

%

BinaryAI achieved higher \texttt{Recall@1} and \texttt{Recall@5} compared to MRT-OAST (see \autoref{table:acc}).
Notably, the \texttt{Recall@1} and \texttt{Recall@5} of BinaryAI showed little difference. This is because the tool was not originally designed to identify similar binary functions for a given source input, and even after our adaptation, it often produced fewer than five candidate binaries per source function.
Overall, the accuracy of BinaryAI remained consistent, except when the \texttt{-O2} optimization was applied.


\vspace{2px}
\noindent\textbf{\textit{Result analysis: \sys.}}
\sys outperformed existing approaches
by achieving 0.7513 and 0.8098 
\texttt{Recall@1} and \texttt{Recall@5}, respectively.
%

Despite its effectiveness, \sys occasionally produced false results due to inaccurate decompilation or extreme code transformations. Although it substantially mitigates the effects of inlining, its accuracy under the \texttt{-O2} optimization slightly decreased because of extreme code logic modifications. However, the \texttt{Recall@5} drop rate of \sys from \texttt{-O0} to \texttt{-O2} was 14.99\%, which is considerably smaller than that of MRT-OAST (44.31\%) and BinaryAI (31.09\%).
In addition, because \sys incorporates syntactic features in block comparison, its accuracy was slightly lower on stripped binaries than on \texttt{no\_strip} binaries. 
Moreover, exceptionally long decompiled functions sometimes caused incorrect mappings even though length-based filtering was applied in P3. 
Nevertheless, \sys achieved the highest accuracy across all configurations, confirming that its design is well-suited for measuring similarity between source and binary functions.

\begin{tcolorbox}[colback=blue3!30, 
    colframe=black, 
    boxrule=0pt, 
    left=5pt,
    right=5pt,
    top=2pt,
    bottom=2pt,
    enhanced jigsaw, 
    sharp corners]
    \textbf{Finding 1.} \sys outperformed the MRT-OAST and BinaryAI by achieving higher \texttt{Recall@1} and \texttt{Recall@5} in all configurations. 
    The block-based function similarity matching of \sys achieved substantially higher detection accuracy than existing approaches, even under \texttt{strip} and \texttt{-O2} optimization settings.
    %
\end{tcolorbox}

\begin{figure}[t]
	\begin{center}
         \begin{subfigure}[b]{0.48\textwidth}
			\centering
                \includegraphics[width=\linewidth]{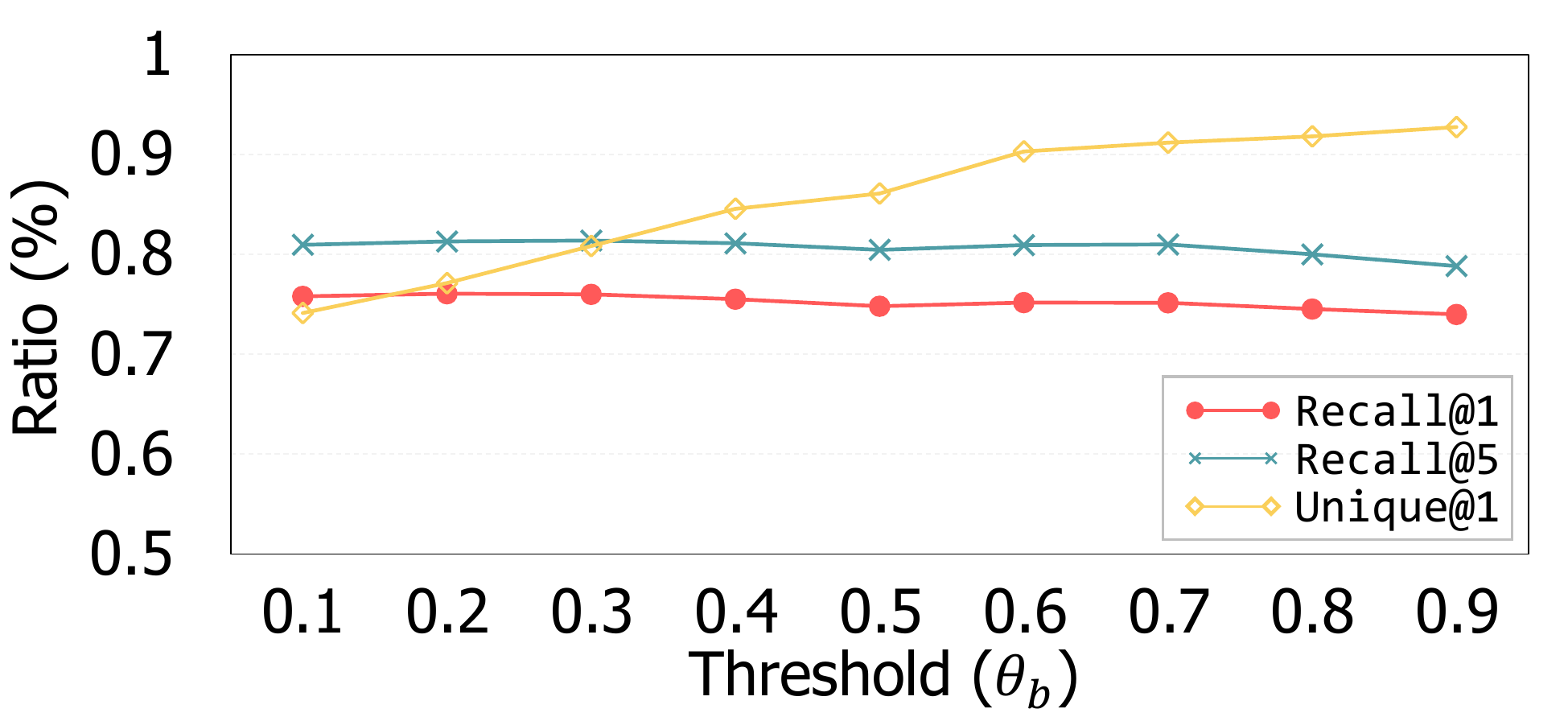}
			
			\caption{Experimental results for $\theta_b$ sensitivity. \texttt{Unique@1} represents the proportion of cases where the top-1 result is uniquely ranked without ties.}
			\label{fig:theta}
            \vspace{0.5em}
		\end{subfigure}\hspace{0.75em}%
		\begin{subfigure}[b]{0.48\textwidth}
			\centering
                \includegraphics[width=\linewidth]{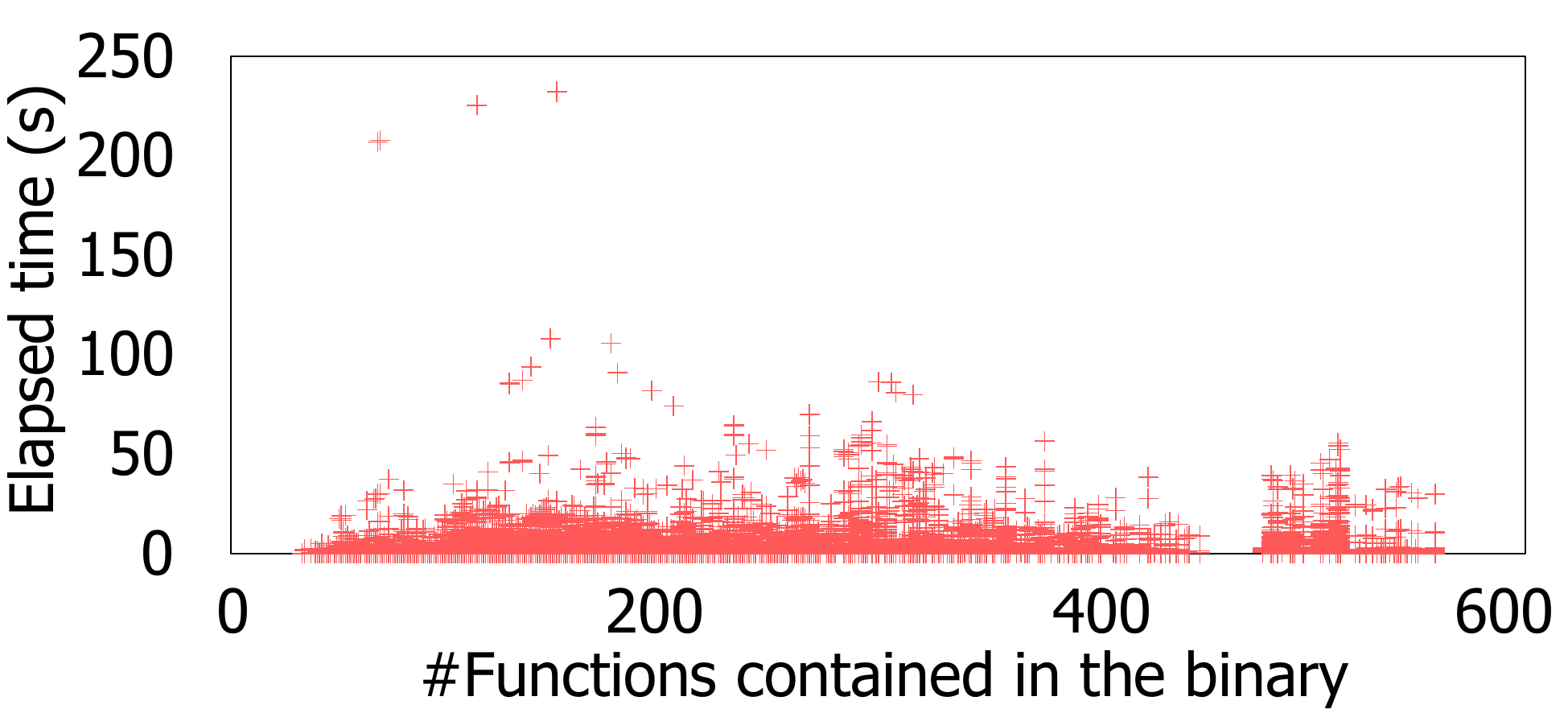}
			
			\caption{Matching time per source function according to the number of binary function candidates. 96.8\% of the cases were matched in under 10 s on average.}
			\label{fig:performance}
            \vspace{0.5em}
		\end{subfigure}

        \vspace{-1em}
		\caption{Threshold experiment and performance evaluation results.}
	\end{center}
	
	\vspace{-1.3em}
\end{figure}

\PP{Threshold sensitivity.}
To measure threshold sensitivity, we varied $\theta_b$ from 0.1 to 0.9 in increments of 0.1 and evaluated \sys's recalls. 
Moreover, as $\theta$ decreases, the matching becomes looser, which may cause two unrelated blocks to be determined as similar. To analyze this effect, we introduce a metric called \texttt{Unique@1}, which represents the proportion of cases where exactly one binary function is identified as the most similar to a given source function.
\autoref{fig:theta} presents the experimental results.
Notably, \sys's accuracy did not vary significantly with different values of $\theta_b$.
However, when the threshold was set too high (\eg, 0.9), similar blocks could not be detected, leading to a slight decrease in recall, whereas when it was set too low (\eg, below 0.6), unrelated blocks were identified as similar, resulting in a decrease in \texttt{Unique@1}.
Therefore, we set $\theta_b$ to 0.7 to achieve high recall while effectively distinguishing similar functions from non-similar ones.


\begin{table}[t]
\renewcommand{\tabcolsep}{2mm}
\caption{\label{table:inlinesb}Accuracy of \sys in inlined function detection.}
\vspace{-1em}
\small
\begin{center}
\begin{tabular}{|c|c|c|c|c|c|c|}
\hline
\rule{0in}{2.2ex}\textbf{Testset} & \textbf{Compiler} & \textbf{Optimization} & \textbf{Inline} $^\dagger$(\%) & \textbf{Symbol}. & \textbf{\texttt{Recall@1}} & \textbf{\texttt{Recall@5}} \\\hline\hline
\rule{0in}{2.2ex}\multirow{4}{*}{\begin{tabular}[c]{@{}c@{}}Coreutils\end{tabular}} & \texttt{Clang} & \texttt{O2} & 54.12\% & \texttt{no\_strip} & 0.5981 & 0.7315 \\
 & \texttt{Clang} & \texttt{O2} & 54.12\% & \texttt{strip} & 0.5624 & 0.6712 \\
 & \texttt{GCC} & \texttt{O2} & 45.65\% & \texttt{no\_strip} & 0.5930 & 0.7343 \\
 & \texttt{GCC} & \texttt{O2} & 45.65\% & \texttt{strip} & 0.5284 & 0.6595 \\\hline\hline
\multirow{4}{*}{\begin{tabular}[c]{@{}c@{}}Inetutils\end{tabular}} & \rule{0in}{2.2ex}\texttt{Clang}& \texttt{O2} & 11.33\% & \texttt{no\_strip} & 0.7297 & 0.7613 \\
 & \texttt{Clang} & \texttt{O2} & 11.33\% & \texttt{strip} & 0.6441 & 0.7162 \\
 & \texttt{GCC} & \texttt{O2} & 9.29\% & \texttt{no\_strip} & 0.5879 & 0.6593 \\
 & \texttt{GCC} & \texttt{O2} & 9.29\% & \texttt{strip} & 0.6319 & 0.6593 \\\hline
\end{tabular}
\end{center}

\vspace{0.1em}
\footnotesize{$\dagger$: The proportion of inlined functions among all binary functions.}

\end{table}

\subsection{Efficacy of \sys} \label{subsec:rq2}
Next, we evaluate how effectively \sys handles function inlining.
Within the experimental results of \autoref{subsec:rq1}, we 
considered only inlined functions and 
examined how well \sys identified them.
\autoref{table:inlinesb} presents the measurement results.
For Coreutils and Inetutils, we observed that when compiled with \texttt{O2} optimization, approximately 50\% and 10\% of the functions, respectively, were inlined.
This is because function inlining decisions depend on various factors (\eg, function size and call frequency) and occur in an unpredictable manner.

Nevertheless, \sys successfully identified the corresponding binary functions for input source functions in binaries where both function inlining and stripping were applied, achieving a \texttt{Recall@1} of at least 0.5284 (up to 0.6441) and a \texttt{Recall@5} of at least 0.6593 (up to 0.7613). When the binary is not stripped, the maximum \texttt{Recall@1} and \texttt{Recall@5} increase to 0.7297 and 0.7613. Existing approaches have not addressed function inlining in function mapping. For example, BinaryAI achieved a \texttt{Recall@1} of less than 0.01 (\ie, 1\%) when applied solely to inlined functions. Therefore, this result strongly demonstrates \sys's effectiveness in identifying inlined functions.




\begin{tcolorbox}[colback=blue3!30, 
    colframe=black, 
    boxrule=0pt, 
    left=5pt,
    right=5pt,
    top=2pt,
    bottom=2pt,
    enhanced jigsaw, 
    sharp corners]
    \textbf{Finding 2.} \sys, which utilizes block-based function matching and a length-based weighting mechanism, can identify inlined functions with achieving up to 0.7297 \texttt{Recall@1} and 0.7613 \texttt{Recall@5}, whereas existing approaches struggle to detect them (\eg, less than 0.01 \texttt{Recall@1}).
\end{tcolorbox}


\subsection{Performance of \sys} \label{subsec:rq3}
To evaluate the performance and scalability of \sys, we measured the time required to identify binary functions similar to the corresponding input source code using 3,904 binaries from the accuracy measurement experiment (see \autoref{subsec:rq1}).
We excluded preprocessing steps that use external libraries (\eg, the time taken by the Joern parser) and considered only the time required to identify similar functions, as preprocessing time depends on external tools and their implementations.
We can measure the execution time either based on the binary size or on the number of functions contained in the binary. However, the former is highly variable due to factors such as optimization or debugging information (\ie, external influences). Therefore, we measured the execution time of \sys based on the latter (\ie, the number of functions),
as it directly determines the number of candidate comparisons and thus dominates the overall matching cost.

\autoref{fig:performance} illustrates the measurement result. When comparing a single input source function with a single binary (with multiple functions), \sys took an average of 1.9 s (median 0.692 s). For 96.8\% of the input functions, \sys took less than 10 s, which demonstrates that \sys is sufficiently fast for practical use.
Next, even as the number of functions in the binary increased, the mapping time did not increase significantly. Therefore, unless a block contains an extreme amount of block content (\eg, cryptography functions), the performance of \sys remains fast. This demonstrates that \sys can operate efficiently and scale well even for large binaries.


\begin{tcolorbox}[colback=blue3!30, 
    colframe=black, 
    boxrule=0pt, 
    left=5pt,
    right=5pt,
    top=2pt,
    bottom=2pt,
    enhanced jigsaw, 
    sharp corners]
    \textbf{Finding 3.} With an average identification time of 1.9 s to detect similar functions in the target binary, \sys demonstrates performance that is sufficient for practical use.
\end{tcolorbox}



\subsection{Application of \sys} \label{subsec:rq4}
In this section, we applied \sys to 1-day vulnerability detection and evaluated its effectiveness.


\PP{Methodology}
We compared \sys with \textsc{React}~\cite{zhan2024react}, a state-of-the-art approach for checking whether a vulnerability patch has been applied in binaries. 
\textsc{React} identifies the target binary function based on the name of the vulnerable function, and then verifies the patch application using intermediate representations (IR). 
We apply \sys to vulnerability detection by first locating the binary function most similar to the vulnerable function.
For a fair comparison, if the function name is identifiable, we utilize this information.
\sys then focuses on the blocks in the vulnerable (\textit{resp}. patched) function that contain deleted (\textit{resp}. added) lines from the patch~\cite{woo2022movery, woo2023v1scan}. After mapping these blocks to all blocks of the target binary function, \sys computes the highest similarity scores for the vulnerable and patched blocks, denoted as $\alpha$ and $\beta$, respectively. If $\alpha \geq\beta$, \sys determines that the vulnerability is present in the binary.

To evaluate the practical applicability of our approach to vulnerability detection, we compiled both vulnerable and patched binaries for each selected CVE in eight configurations (2 compilers $\times$ 2 optimizations $\times$ 2 symbol management options). 
\textsc{React} was evaluated on four OSS projects: \texttt{LibXML2}, \texttt{tcpdump}, \texttt{OpenSSL}, and \texttt{FFmpeg}.
We utilized all CVEs for \texttt{LibXML2} and \texttt{tcpdump} from the \textsc{React}'s dataset, as their binaries are relatively small and contain a manageable number of functions. In contrast, \texttt{OpenSSL} and \texttt{FFmpeg} presented a significant challenge due to their large scale and high variance between patched versions. Therefore, we selected only CVEs whose patched binaries belonged to the latest major version within \textsc{React}'s dataset. This process resulted in 416 binaries from 26 CVEs (208 vulnerable and 208 patched).  
We determined that this dataset is sufficient for our purposes, because the primary goal of this experiment is to show that \sys, our tool for detecting similar functions, can effectively be extended to vulnerability detection.

Detection results are classified as follows: correctly identifying a vulnerable function in vulnerable binaries (TP), failing to detect it (FN), incorrectly flagging a patched binary as vulnerable (FP), and correctly identifying a patched binary as safe (TN).

We evaluated \sys and \textsc{React} based on the following three metrics: \texttt{precision} (P), \texttt{recall} (R), and \texttt{F1} \texttt{score} (F1).

\begin{table}[t]
    \renewcommand{\tabcolsep}{0.85mm}
    \caption{\label{table:vuln}Vulnerability detection results (\sys vs. \textsc{React}).}
    \footnotesize
    \begin{center}
    \begin{tabular}{|c|c|r|r|r|r|r|r|r|r|r|r|r|r|r|r|}
    \hline
    \rule{0in}{2.2ex}\multirow{2}{*}{\textbf{OSS}} &\multirow{2}{*}{\textbf{\#CVEs}} 
    &\multicolumn{7}{c|}{\textbf{\textsc{React}}}
    &\multicolumn{7}{c|}{\textbf{\sys}}\\\cline{3-16}
    & &  
    \multicolumn{1}{c|}{\rule{0in}{2.2ex}\texttt{\textbf{TP}}} 
    & \multicolumn{1}{c|}{\texttt{\textbf{FP}}} 
    & \multicolumn{1}{c|}{\texttt{\textbf{TN}}} 
    & \multicolumn{1}{c|}{\texttt{\textbf{FN}}}
    & \multicolumn{1}{c|}{\texttt{\textbf{Precision}}}
    & \multicolumn{1}{c|}{\texttt{\textbf{Recall}}}
    & \multicolumn{1}{c|}{\texttt{\textbf{F1}} \texttt{\textbf{score}}}
    & \multicolumn{1}{c|}{\texttt{\textbf{TP}}} 
    & \multicolumn{1}{c|}{\texttt{\textbf{FP}}} 
    & \multicolumn{1}{c|}{\texttt{\textbf{TN}}} 
    & \multicolumn{1}{c|}{\texttt{\textbf{FN}}}
    & \multicolumn{1}{c|}{\texttt{\textbf{Precision}}}
    & \multicolumn{1}{c|}{\texttt{\textbf{Recall}}}
    & \multicolumn{1}{c|}{\texttt{\textbf{F1}} \texttt{\textbf{score}}}\\\hline\hline
    \rule{0in}{2.2ex}\texttt{tcpdump}
    & 10
    & 34 & 20 & 14 & 92 & 0.6296 & 0.2698 & 0.3778 
    & 55 & 23 & 39 & 43 & 0.7051
    & 0.5612 & 0.6250\\
    \texttt{OpenSSL}
    & 8
    & 32 & 7 & 25 & 64 & 0.8205 & 0.3333 & 0.4741 
    & 30 & 22 & 11 & 65 & 0.5769
    & 0.3158 & 0.4082\\
    \texttt{LibXML2}
    & 2
    & 8 & 4 & 4 & 16 & 0.6667 & 0.3333 & 0.4444 
    & 13 & 6 & 6 & 7 & 0.6842
    & 0.6500 & 0.6667\\
    \texttt{FFmpeg}
    & 6
    & 20 & 4 & 18 & 54 & 0.8333 & 0.2703 & 0.4082 
    & 18 & 28 & 6 & 44 & 0.3913
    & 0.2903 & 0.3333 \\\hline\hline
    \multicolumn{1}{|c|}{\rule{0in}{2.2ex}Total} & 26 & 94 & \textcolor{tomato}{35} & 61 & 226 & \textcolor{tomato}{0.7287} & 0.2938 & 0.4187 & \textcolor{tomato}{116} & 79 & \textcolor{tomato}{62} & \textcolor{tomato}{159} & 0.5949 & \textcolor{tomato}{0.4218} & \textcolor{tomato}{0.4936} \\\hline
    
    \end{tabular}
    \end{center}

\end{table}

\vspace{0.5em}
\begin{center}
    $P = \dfrac{\text{\texttt{\#TP}}}{\text{\texttt{\#TP}} + \text{\texttt{\#FP}}}$,\hspace{1em}
    $R = \dfrac{\text{\texttt{\#TP}}}{\text{\texttt{\#TP}} + \text{\texttt{\#FN}}}$,\hspace{1em}
    $\text{\texttt{F1}} = \dfrac{2* \text{\texttt{P}} * \text{\texttt{R}}}{\text{\texttt{P}} + \text{\texttt{R}}}$
\end{center}
\vspace{0.5em}

\PP{Result analysis} 
Despite applying \sys to vulnerability detection in an intuitive manner, \sys demonstrated slightly higher accuracy (\ie, \texttt{F1} \texttt{score}) than \textsc{React}.
Notably, \sys identified more TPs and fewer FNs than \textsc{React}.
The primary cause of the high FN rate in \textsc{React} was its reliance on function names, which made it nearly infeasible to test the presence of patches in stripped binaries. Furthermore, when function inlining occurred under O2 optimization, \textsc{React}'s accuracy dropped significantly.
In contrast, \sys could trace vulnerable functions even in O2-optimized or stripped binaries, yielding fewer FNs. However, because code similarity–based vulnerability detection is less effective when only small portions of the vulnerable code are modified in the patch~\cite{kim2017vuddy, woo2022movery}, it produced many FPs.
Identifying 1-day vulnerabilities in stripped or O2-optimized binaries based on vulnerable source functions remains a challenging open problem. Our attempt to apply \sys to this task 
achieved reasonable accuracy and demonstrates the potential of this approach as a new direction for vulnerability detection.

\begin{tcolorbox}[colback=blue3!30, 
    colframe=black, 
    boxrule=0pt, 
    left=5pt,
    right=5pt,
    top=2pt,
    bottom=2pt,
    enhanced jigsaw, 
    sharp corners]
    \textbf{Finding 4.} \sys's technique for detecting similar functions across domains can also be effectively utilized for identifying propagated vulnerabilities.
\end{tcolorbox}




\vspace{0.5px}
\section{Discussion}\label{sec:dis}


\PP{Addressing function inlining}
\sys addresses function inlining through two main ideas: (1) leveraging source code as a reference and (2) applying a length-based weighting mechanism. When computing function similarity (see \autoref{subsec:p3}), the reference point is the total number of blocks in the source function. Thus, as long as a sufficient portion of the source blocks is preserved in the binary function, it can still be identified as a similar match.
Moreover, by incorporating a length-based weighting mechanism, \sys remains robust even when inlining increases the size of the binary function compared to the source function. Together, these strategies enable \sys to effectively handle common inlining scenarios.

\PP{String block ablation.} Although string literals are among the most robust features across compilation, they do not directly belong to control flow. To evaluate string block-specific impact, we conducted an ablation study that excludes string blocks. The results showed only a minimal recall drop of no more than 3\%, indicating that string blocks are not critical to matching accuracy, but instead provide a complementary benefit.


\PP{Limitations and future work}
First, 
%
\sys depends on the quality of decompilation. If the decompiler fails to accurately reconstruct source-level semantics, especially under aggressive compiler optimizations, the resulting mismatches can lead to false mappings.
Next,
although \sys mitigates some effects of compiler optimizations and symbol removal, challenges remain in handling function inlining and stripped binaries. These factors still cause slight accuracy degradation. We plan to incorporate semantic features beyond syntactic block comparison (\eg, data-flow) to improve robustness in such environments.
Lastly, 
we only considered \texttt{O0} and \texttt{O2} optimizations. We assumed that \texttt{O1} would be covered by \texttt{O2} and chose \texttt{O2} over \texttt{O3} because it is more widely used and better suited to illustrating cases where function inlining and syntax transformations significantly differ. If this becomes a concern, we will extend our evaluation to include \texttt{O1} and \texttt{O3} optimizations. 

\PP{Threats to validity}
To demonstrate the generality of \sys, we selected the two most binary-rich datasets from BinKit. However, these may not fully represent the entire software ecosystem. 
Next, due to the absence of ground truth, we created our own ground truth under the assumption that \texttt{O0} binaries are not inlined (see \autoref{subsec:rq1}). This assumption may have a minor impact on the accuracy measurement.
In addition, we adapted MRT-OAST and BinaryAI for our purposes to conduct comparative experiments. While we aimed to ensure fairness in our experiments, there may be cases where existing approaches' accuracy might not be correctly evaluated. 
Our goal is not to undermine them but to demonstrate that \sys effectively identifies binary functions similar to a given source code.
%
Next,
although \sys detected similar binary functions for the input function in an average of 1.9 s (see \autoref{subsec:rq3}), this time may vary depending on the environment. 
Finally, although \sys can utilize any decompilation tool, we used Ghidra in our experiments. If a different tool is used, the accuracy may vary slightly. 

\section{Related Work}



\vspace{-4px}
\PP{Source-to-binary matching}
Existing source-to-binary approaches (\eg, \cite{yuan2019b2sfinder, duan2017identifying, ban2021b2smatcher, jiang2024binaryai, dong2024libvdiff}) have 
primarily focused on identifying reused OSS components in binaries by leveraging compilation-resilient features (\eg, string literals). 
ISRD~\cite{xu2021interpretation}, LibDB~\cite{Tang2022libDB}, LibAM~\cite{li2023libam}, and ModX~\cite{yang2022modx} used function-level similarity features to identify OSS components. LibvDiff~\cite{dong2024libvdiff} focuses on OSS version identification in binaries by combining source-level differences with function-level binary similarity.
REACT~\cite{zhan2024react} verified patch presence by employing symbolic execution on IR code from both source and binary.
BinaryAI~\cite{jiang2024binaryai} attempted to identify reused OSS components by measuring code similarity between the embeddings of decompiled binary code and source functions.  CodeCMR~\cite{codecmr} addresses the representation disparity by using node embeddings with DPCNN and GNN. CrossCode2Vec~\cite{crosscode2vec2025} addresses the compilation gap through unified embeddings across source code and its corresponding compiled binary.
However, existing techniques either take a coarse-grained approach or rely on limited features, making them unsuitable for analyzing 
similarities between source and binary functions. Even the recent AI-based method fails to adequately handle function inlining, rendering it ineffective for our target problem.

\vspace{-1px}
\PP{Binary-to-binary matching}
Existing binary-to-binary matching approaches aimed to assess structural or syntactic similarity by leveraging text hashing of assembly code~\cite{gitz2017, viva}, matching control flow graphs~\cite{bourquin2013binslayer}, or performing symbolic execution~\cite{gao2008binhunt,zhan2024ps3,ming2017binsim}.
However, these approaches face scalability limitations due to obfuscation, compiler optimizations, and path explosion issues.
%
Alternatively, deep learning models, including Transformers and graph neural networks (GNNs), have been introduced, improving the performance of binary code similarity measurement~\cite{he2024code, wang2022jtrans, zhu2023ktrans, massarelli2019safe, liu2018alphadiff,xu2023improving, he2024bingo}. 
Recent approaches, such as Binshot~\cite{ahn2022practical} and OrderMatters~\cite{yu2020order}, employ pre-trained BERT models. BinXray~\cite{xu2020patch} identifies patches by extracting signatures through comparison of vulnerable and patched functions.
However, these approaches are also difficult to apply effectively to our target problem: they (1) are not applicable to bridge the representation gap between binaries and source code and (2) hardly consider function inlining.


\section{Conclusion}
Identifying reused source code in binaries is a pressing issue, but the significant gap between source and binary representations makes it highly challenging. In response, we present \sys, an approach that effectively identifies binary functions similar to given source functions. By leveraging control block-based function matching, 
\sys outperforms existing approaches in mapping source and binary functions. This enables \sys to remain effective even when direct source-to-binary correspondence is obscured by compiler transformations, such as optimization and function inlining. With \sys, reused source code in binaries can be identified with high accuracy and scalability, achieving 75.13\% Recall@1 and 80.98\% Recall@5, with an average matching time of 1.9 seconds per source function. Furthermore, as our experiments demonstrate, \sys can also be applied to detect propagated vulnerabilities, contributing to a safer software ecosystem.

\section*{Data Availability}
\sys is available at \url{https://github.com/heedongy/SBridge_Artifact}.

\begin{acks}
This work was supported by the Institute of Information\& Communications Technology Planning \& Evaluation (IITP) grant funded by the Korea government (MSIT) (RS-2024-00440780, Development of Automated SBOM and VEX Verification Technologies for Securing Software Supply Chains), ICT Creative Consilience Program (IITP-2026-RS-2020-II201819), the National Research Foundation of Korea (NRF) grant funded by the Korea government (MSIT) (RS-2025-00517788, Research on Intelligent SBOM Generation and Automated Vulnerability Analysis through Multi-level Code Analysis) and the Culture, Sports and Tourism R\&D Program through the Korea Creative Content Agency grant funded by the Ministry of Culture, Sports and Tourism (International Collaborative Research and Global Talent Development for the Development of Copyright Management and Protection Technologies for Generative AI, RS-2024-00345025).

\end{acks}

\bibliographystyle{ACM-Reference-Format}
\bibliography{fse_ref}

@inproceedings{woo2021centris,
	title={{CENTRIS: A Precise and Scalable Approach for Identifying Modified Open-Source Software Reuse}},
	author={Woo, Seunghoon and Park, Sunghan and Kim, Seulbae and Lee, Heejo and Oh, Hakjoo},
	booktitle={Proceedings of the IEEE/ACM 43rd International Conference on Software Engineering (ICSE)},
    doi = {10.1109/ICSE43902.2021.00083},
	pages={860--872},
	year={2021}
}

@inproceedings{yuan2019b2sfinder,
  title={{B2SFinder: Detecting Open-Source Software Reuse in COTS Software}},
  author={Feng, Muyue and Yuan, Zimu and Li, Feng and Ban, Gu and Xiao, Yang and Wang, Shiyang and Tang, Qian and Su, He and Yu, Chendong and Xu, Jiahuan and Piao, Aihua and Xue, Jingling and Huo, Wei},
  year = {2020},
  isbn = {9781728125084},
  publisher = {IEEE Press},
  url = {https://doi.org/10.1109/ASE.2019.00100},
  doi = {10.1109/ASE.2019.00100},
  booktitle = {Proceedings of the 34th IEEE/ACM International Conference on Automated Software Engineering},
  pages = {1038–1049},
  numpages = {12},
  location = {San Diego, California},
  series = {ASE '19}
}

@Misc{ctags,
	title = 	{{Universal Ctags}},
	author = {Ctags},
	year = 2024,
	note = {\url{https://github.com/universal-ctags/ctags}}
}

@inproceedings{yamaguchi2014modeling,
	title={{Modeling and Discovering Vulnerabilities with Code Property Graphs}},
	author={Yamaguchi, Fabian and Golde, Nico and Arp, Daniel and Rieck, Konrad},
	booktitle={Proceedings of the 35th IEEE Symposium on Security and Privacy (SP)},
	pages={590--604},
	year={2014},
    doi={10.1109/SP.2014.44},
	organization={IEEE}
}

@inproceedings{xu2020patch,
	title={{Patch Based Vulnerability Matching for Binary Programs}},
	author={Xu, Yifei and Xu, Zhengzi and Chen, Bihuan and Song, Fu and Liu, Yang and Liu, Ting},
	booktitle={Proceedings of the 29th ACM SIGSOFT International Symposium on Software Testing and Analysis (ISSTA)},
	pages={376--387},
    doi = {10.1145/3395363.3397361},
	year={2020}
}

@inproceedings{woo2022movery,
	title={{MOVERY: A Precise Approach for Modified Vulnerable Code Clone Discovery from Modified Open-Source Software Components}},
	author={Woo, Seunghoon and Hong, Hyunji and Choi, Eunjin and Lee, Heejo},
	booktitle={Proceedings of the 31st USENIX Security Symposium (Security)},
	pages={3037--3053},
	year={2022}
}

@inproceedings{woo2023v1scan,
  title={{V1SCAN: Discovering 1-day Vulnerabilities in Reused C/C++ Open-source Software Components Using Code Classification Techniques}},
  author={Woo, Seunghoon and Choi, Eunjin and Lee, Heejo and Oh, Hakjoo},
  booktitle={Proceedings of the 32nd USENIX Security Symposium (Security)},
  pages={6541--6556},
  year={2023}
}

@inproceedings{woo2021v0finder,
	title={{V0Finder: Discovering the Correct Origin of Publicly Reported Software Vulnerabilities}},
	author={Woo, Seunghoon and Lee, Dongwook and Park, Sunghan and Lee, Heejo and Dietrich, Sven},
	booktitle={Proceedings of the 30th USENIX Security Symposium (Security)},
	pages={3041--3058},
	year={2021}
}

@inproceedings{xiao2020mvp,
	title={{MVP: detecting vulnerabilities using patch-enhanced vulnerability signatures}},
	author={Xiao, Yang and Chen, Bihuan and Yu, Chendong and Xu, Zhengzi and Yuan, Zimu and Li, Feng and Liu, Binghong and Liu, Yang and Huo, Wei and Zou, Wei and Shi, Wenchang},
	booktitle={Proceedings of the 29th USENIX Security Symposium (Security)},
	pages={1165--1182},
	year={2020}
}

@Misc{openssl,
	title = {{A Cryptography and SSL/TLS Toolkit}},
	author = {OpenSSL},
	year = 2021,
	note = {\url{https://www.openssl.org/}}
}

@Misc{ffmpeg,
	title = {{A complete, cross-platform solution to record, convert and stream audio and video.}},
	author = 	{FFmpeg},
	year =			2021,
	note = {\url{https://ffmpeg.org/}}
}

@inproceedings{duan2017identifying,
author = {Duan, Ruian and Bijlani, Ashish and Xu, Meng and Kim, Taesoo and Lee, Wenke},
title = {{Identifying Open-Source License Violation and 1-day Security Risk at Large Scale}},
year = {2017},
isbn = {9781450349468},
publisher = {Association for Computing Machinery},
address = {New York, NY, USA},
url = {https://doi.org/10.1145/3133956.3134048},
doi = {10.1145/3133956.3134048},
booktitle = {Proceedings of the 2017 ACM SIGSAC Conference on Computer and Communications Security},
pages = {2169–2185},
numpages = {17},
location = {Dallas, Texas, USA},
series = {CCS '17}
}

@inproceedings{sajnani2016sourcerercc,
	title={{SourcererCC: Scaling Code Clone Detection to Big-Code}},
	author={Sajnani, Hitesh and Saini, Vaibhav and Svajlenko, Jeffrey and Roy, Chanchal K and Lopes, Cristina V},
	booktitle={Proceedings of the 38th International Conference on Software Engineering (ICSE)},
    doi = {10.1145/2884781.2884877},
	pages={1157--1168},
	year={2016},
}

@inproceedings{wang2018ccaligner,
	title={{CCAligner: A Token Based Large-Gap Clone Detector}},
	author={Wang, Pengcheng and Svajlenko, Jeffrey and Wu, Yanzhao and Xu, Yun and Roy, Chanchal K},
	booktitle={Proceedings of the 40th International Conference on Software Engineering (ICSE)},
    doi = {10.1145/3180155.3180179},
	pages={1066--1077},
	year={2018},
}

@inproceedings{kim2017vuddy,
	title={{VUDDY: A Scalable Approach for Vulnerable Code Clone Discovery}},
	author={Kim, Seulbae and Woo, Seunghoon and Lee, Heejo and Oh, Hakjoo},
	booktitle={Proceedings of the 38th IEEE Symposium on Security and Privacy (SP)},
	pages={595--614},
    doi={10.1109/SP.2017.62},
	year={2017},
}

@inproceedings{jiang2024binaryai,
author = {Jiang, Ling and An, Junwen and Huang, Huihui and Tang, Qiyi and Nie, Sen and Wu, Shi and Zhang, Yuqun},
title = {{BinaryAI: Binary Software Composition Analysis via Intelligent Binary Source Code Matching}},
year = {2024},
isbn = {9798400702174},
publisher = {Association for Computing Machinery},
address = {New York, NY, USA},
url = {https://doi.org/10.1145/3597503.3639100},
doi = {10.1145/3597503.3639100},
booktitle = {Proceedings of the IEEE/ACM 46th International Conference on Software Engineering},
articleno = {224},
numpages = {13},
location = {Lisbon, Portugal},
series = {ICSE '24}
}

@inproceedings{dong2024libvdiff,
	title={{LibvDiff: Library Version Difference Guided OSS Version Identification in Binaries}},
	author={Dong, Chaopeng and Li, Siyuan and Yang, Shougou and Xiao, Yang and Wang, Yongpan and Li, Hong and Li, Zhi and Sun, Limin},
	booktitle={Proceedings of the 46th International Conference on Software Engineering (ICSE)},
    doi = {10.1145/3597503.3623336},
	pages={791--802},
	year={2024},
}

@inproceedings{na2024cneps,
	title={{CNEPS: A Precise Approach for Examining Dependencies Among Third-Party C/C++ Open-Source Components}},
	author={Na, Yoonjong and Woo, Seunghoon and Lee, Joomyeong and Lee, Heejo},
	booktitle={Proceedings of the 46th International Conference on Software Engineering (ICSE)},
    doi = {10.1145/3597503.3639209},
	pages={2918--2929},
	year={2024},
}

@inproceedings{Tang2022libDB,
author = {Tang, Wei and Wang, Yanlin and Zhang, Hongyu and Han, Shi and Luo, Ping and Zhang, Dongmei},
title = {{LibDB: an effective and efficient framework for detecting third-party libraries in binaries}},
year = {2022},
isbn = {9781450393034},
publisher = {Association for Computing Machinery},
address = {New York, NY, USA},
url = {https://doi.org/10.1145/3524842.3528442},
doi = {10.1145/3524842.3528442},
booktitle = {Proceedings of the 19th International Conference on Mining Software Repositories},
pages = {423–434},
numpages = {12},
location = {Pittsburgh, Pennsylvania},
series = {MSR '22}
}

@INPROCEEDINGS{xu2021interpretation,
  author = {Xu, Xi and Zheng, Qinghua and Yan, Zheng and Fan, Ming and Jia, Ang and Liu, Ting},
  booktitle={Proceedings of the 43rd International Conference on Software Engineering (ICSE)},
  doi={10.1109/ICSE43902.2021.00084},
  title={{Interpretation-enabled Software Reuse Detection Based on a Multi-Level Birthmark Model}},
  year={2021}
}

@Misc{idapro,
	title = {{IDA Pro}},
	author = {Hex-Rays},
	year = 2024,
	note = {\url{https://hex-rays.com/ida-pro/}}
}

@Misc{binaryninja,
	title = 	{{Binary Ninja}},
	author = {Vector 35},
	year = 2024,
	note = {\url{https://binary.ninja/}}
}

@Misc{ghidra,
	title = 	{{Ghidra}},
	author = {National Security Agency},
	year = 2024,
	note = {\url{https://ghidra-sre.org}}
}

@inproceedings{yang2022modx,
title={{ModX: binary level partially imported third-party library detection via program modularization and semantic matching}},
author={Yang, Can and Xu, Zhengzi and Chen, Hongxu and Liu, Yang and Gong, Xiaorui and Liu, Baoxu},
booktitle={Proceedings of the 44th International Conference on Software Engineering},
pages={1393--1405},
doi = {10.1145/3510003.3510627},
year={2022}
}

@inproceedings{tang2020libdx,
title={{LibDX: A Cross-Platform and Accurate System to Detect Third-Party Libraries in Binary Code}},
author={Tang, Wei and Luo, Ping and Fu, Jialiang and Zhang, Dan},
booktitle={2020 IEEE 27th International Conference on Software Analysis, Evolution and Reengineering (SANER)},
pages={104--115},  
doi={10.1109/SANER48275.2020.9054845},
year={2020}
}

@article{li2023libam,
author = {Li, Siyuan and Wang, Yongpan and Dong, Chaopeng and Yang, Shouguo and Li, Hong and Sun, Hao and Lang, Zhe and Chen, Zuxin and Wang, Weijie and Zhu, Hongsong and Sun, Limin},
title = {{LibAM: An Area Matching Framework for Detecting Third-Party Libraries in Binaries}},
year = {2023},
publisher = {Association for Computing Machinery},
address = {New York, NY, USA},
issn = {1049-331X},
doi = {10.1145/3625294},
journal = {ACM Trans. Softw. Eng. Methodol.},
month = {sep}
}

@inproceedings{revng,
  title={{rev.ng: a unified binary analysis framework to recover CFGs and function boundaries}},
  author={Di Federico, Alessandro and Payer, Mathias and Agosta, Giovanni},
  booktitle={Proceedings of the 26th International Conference on Compiler Construction},
  pages={131--141},
  year={2017}
}

@ARTICLE{binkit,
  author={Kim, Dongkwan and Kim, Eunsoo and Cha, Sang Kil and Son, Sooel and Kim, Yongdae},
  journal={IEEE Transactions on Software Engineering},
  title={{Revisiting Binary Code Similarity Analysis Using Interpretable Feature Engineering and Lessons Learned}},
  year={2023},
  volume={49},
  number={4},
  pages={1661-1682},
  doi={10.1109/TSE.2022.3187689}
}

@Manual{blackduck2025report,
	title = {{2025 Open Source Security and Risk Analysis Report}},
	organization = {Synopsys},
	year = 2025
}

@article{woo2025large,
title = {{A large-scale analysis of the effectiveness of publicly reported security patches}},
journal = {Computers \& Security},
volume = {148},
pages = {104181},
year = {2025},
issn = {0167-4048},
doi = {https://doi.org/10.1016/j.cose.2024.104181},
url = {https://www.sciencedirect.com/science/article/pii/S0167404824004863},
author = {Seunghoon Woo and Eunjin Choi and Heejo Lee}
}

@inproceedings{jia2024cross,
author = {Jia, Ang and Fan, Ming and Xu, Xi and Jin, Wuxia and Wang, Haijun and Liu, Ting},
title = {{Cross-Inlining Binary Function Similarity Detection}},
year = {2024},
isbn = {9798400702174},
publisher = {Association for Computing Machinery},
address = {New York, NY, USA},
url = {https://doi.org/10.1145/3597503.3639080},
doi = {10.1145/3597503.3639080},
booktitle = {Proceedings of the IEEE/ACM 46th International Conference on Software Engineering},
articleno = {223},
numpages = {13},
location = {Lisbon, Portugal},
series = {ICSE '24}
}

@inproceedings{jiang2023third,
  title={{Third-party library dependency for large-scale sca in the c/c++ ecosystem: How far are we?}},
  author={Jiang, Ling and Yuan, Hengchen and Tang, Qiyi and Nie, Sen and Wu, Shi and Zhang, Yuqun},
  booktitle={Proceedings of the 32nd ACM SIGSOFT International Symposium on Software Testing and Analysis},
  doi = {10.1145/3597926.3598143},
  pages={1383--1395},
  year={2023}
}

@inproceedings{zhan2024react,
  title={{REACT: IR-Level Patch Presence Test for Binary}},
  author={Zhan, Qi and Hu, Xing and Xia, Xin and Li, Shanping},
  booktitle={Proceedings of the 39th IEEE/ACM International Conference on Automated Software Engineering},
  doi = {10.1145/3691620.3695012},
  pages={381--392},
  year={2024}
}

@inproceedings{gitz2017,
  title={{Similarity of binaries through re-optimization}},
  author={David, Yaniv and Partush, Nimrod and Yahav, Eran},
  booktitle={Proceedings of the 38th ACM SIGPLAN conference on programming language design and implementation},
  doi = {10.1145/3140587.3062387},
  pages={79--94},
  year={2017}
}

@inproceedings{viva,
  author={Xiao, Yang and Xu, Zhengzi and Zhang, Weiwei and Yu, Chendong and Liu, Longquan and Zou, Wei and Yuan, Zimu and Liu, Yang and Piao, Aihua and Huo, Wei},
  booktitle={2021 IEEE International Conference on Software Analysis, Evolution and Reengineering (SANER)},
  title={{VIVA: Binary Level Vulnerability Identification via Partial Signature}},
  year={2021},
  pages={213-224},
  doi={10.1109/SANER50967.2021.00028}}

@inproceedings{massarelli2019safe,
  title={{SAFE: Self-Attentive Function Embeddings for Binary Similarity}},
  author={Massarelli, Luca and Di Luna, Giuseppe Antonio and Petroni, Fabio and Baldoni, Roberto and Querzoni, Leonardo},
  booktitle={Detection of Intrusions and Malware, and Vulnerability Assessment: 16th International Conference, DIMVA 2019, Gothenburg, Sweden, June 19--20, 2019, Proceedings 16},
  pages={309--329},
  year={2019},
  doi={10.1007/978-3-030-22038-9_15},
  organization={Springer}
}

@inproceedings{bourquin2013binslayer,
  title={{BinSlayer: Accurate Comparison of Binary Executables}},
  author={Bourquin, Martial and King, Andy and Robbins, Edward},
  booktitle={Proceedings of the 2nd ACM SIGPLAN Program Protection and Reverse Engineering Workshop},
  doi = {10.1145/2430553.2430557},
  pages={1--10},
  year={2013}
}

@inproceedings{liu2018alphadiff,
  title={{$\alpha$Diff: cross-version binary code similarity detection with DNN}},
  author={Liu, Bingchang and Huo, Wei and Zhang, Chao and Li, Wenchao and Li, Feng and Piao, Aihua and Zou, Wei},
  booktitle={Proceedings of the 33rd ACM/IEEE international conference on automated software engineering},
  doi = {10.1145/3238147.3238199},
  pages={667--678},
  year={2018}
}

@inproceedings{wang2022jtrans,
  title={{jTrans: Jump-Aware Transformer for Binary Code Similarity}},
  author={Wang, Hao and Qu, Wenjie and Katz, Gilad and Zhu, Wenyu and Gao, Zeyu and Qiu, Han and Zhuge, Jianwei and Zhang, Chao},
  booktitle={Proceedings of the 31st ACM SIGSOFT International Symposium on Software Testing and Analysis},
  doi = {10.1145/3533767.3534367},
  pages={1--13},
  year={2022}
}

@article{zhu2023ktrans,
  title={{kTrans: Knowledge-Aware Transformer for Binary Code Embedding}},
  author={Zhu, Wenyu and Wang, Hao and Zhou, Yuchen and Wang, Jiaming and Sha, Zihan and Gao, Zeyu and Zhang, Chao},
  journal={arXiv preprint arXiv:2308.12659},
  year={2023}
}

@inproceedings{ahn2022practical,
  title={{Practical Binary Code Similarity Detection with BERT-based Transferable Similarity Learning}},
  author={Ahn, Sunwoo and Ahn, Seonggwan and Koo, Hyungjoon and Paek, Yunheung},
  booktitle={Proceedings of the 38th Annual Computer Security Applications Conference},
  pages={361--374},
  doi = {10.1145/3564625.3567975},
  year={2022}
}

@inproceedings{he2024code,
author = {He, Haojie and Lin, Xingwei and Weng, Ziang and Zhao, Ruijie and Gan, Shuitao and Chen, Libo and Ji, Yuede and Wang, Jiashui and Xue, Zhi},
title = {{Code is not natural language: unlock the power of semantics-oriented graph representation for binary code similarity detection}},
year = {2024},
isbn = {978-1-939133-44-1},
publisher = {USENIX Association},
address = {USA},
booktitle = {Proceedings of the 33rd USENIX Conference on Security Symposium},
articleno = {99},
numpages = {18},
location = {Philadelphia, PA, USA},
series = {SEC '24}
}

@inproceedings{yu2020order,
  title={{Order Matters: Semantic-Aware Neural Networks for Binary Code Similarity Detection}},
  author={Yu, Zeping and Cao, Rui and Tang, Qiyi and Nie, Sen and Huang, Junzhou and Wu, Shi},
  booktitle={Proceedings of the AAAI conference on artificial intelligence},
  volume={34},
  number={01},
  doi={10.1609/aaai.v34i01.5466},
  pages={1145--1152},
  year={2020}
}

@inproceedings{gao2008binhunt,
  title={{BinHunt: Automatically Finding Semantic Differences in Binary Programs}},
  author={Gao, Debin and Reiter, Michael K and Song, Dawn},
  booktitle={International Conference on Information and Communications Security},
  pages={238--255},
  doi = {10.1007/978-3-540-88625-9_16},
  year={2008},
  organization={Springer}
}

@article{ban2021b2smatcher,
  title={{B2SMatcher: fine-Grained version identification of open-Source software in binary files}},
  author={Ban, Gu and Xu, Lili and Xiao, Yang and Li, Xinhua and Yuan, Zimu and Huo, Wei},
  journal={Cybersecurity},
  volume={4},
  pages={1--21},
  year={2021},
  doi={10.1186/s42400-021-00085-7},
  publisher={Springer}
}

@inproceedings{xu2023improving,
  title={{Improving Binary Code Similarity Transformer Models by Semantics-Driven Instruction Deemphasis}},
  doi = {10.1145/3597926.3598121},
  author={Xu, Xiangzhe and Feng, Shiwei and Ye, Yapeng and Shen, Guangyu and Su, Zian and Cheng, Siyuan and Tao, Guanhong and Shi, Qingkai and Zhang, Zhuo and Zhang, Xiangyu},
  booktitle={Proceedings of the 32nd ACM SIGSOFT International Symposium on Software Testing and Analysis},
  pages={1106--1118},
  year={2023}
}

@inproceedings{he2024bingo,
  title={{BinGo: Identifying Security Patches in Binary Code with Graph Representation Learning}},
  author={He, Xu and Wang, Shu and Feng, Pengbin and Wang, Xinda and Sun, Shiyu and Li, Qi and Sun, Kun},
  booktitle={Proceedings of the 19th ACM Asia Conference on Computer and Communications Security},
  doi = {10.1145/3634737.3637666},
  pages={1186--1199},
  year={2024}
}

@manual{gnuclibrary,
	title = {{The GNU C Library Reference Manual, for version 2.42}},
	author = {Sandra Loosemore and Richard M. Stallman and Roland McGrath and Andrew Oram and Ulrich Drepper},
    year = {2025},
    url  = {https://sourceware.org/glibc/manual/2.42/pdf/libc.pdf}
}

@manual{llvmdoxygen,
  title        = {LLVM Project Doxygen Documentation},
  author       = {{LLVM Project}},
  organization = {LLVM Foundation},
  year         = {2025},
  url          = {https://llvm.org/doxygen/}
}

@misc{ibmlibc,
  title        = {{Standard C Library Functions Table, By Name}},
  author       = {{IBM}},
  year         = {2025},
  howpublished = {\url{https://www.ibm.com/docs/en/i/7.6.0?topic=extensions-standard-c-library-functions-table-by-name}}
}

@inproceedings{yu2025multiple,
  title={{A Multiple Representation Transformer with Optimized Abstract Syntax Tree for Efficient Code Clone Detection}},
  author={{Yu, Tianchen and Yuan, Li and Lin, Liannan and He, Hongkui}},
  booktitle={2025 IEEE/ACM 47th International Conference on Software Engineering (ICSE)},
  pages={587--587},
  year={2025},
  doi={10.1109/ICSE55347.2025.00050},
  organization={IEEE Computer Society}
}

@inproceedings{codecmr,
author = {Yu, Zeping and Zheng, Wenxin and Wang, Jiaqi and Tang, Qiyi and Nie, Sen and Wu, Shi},
title = {{CodeCMR: cross-modal retrieval for function-level binary source code matching}},
year = {2020},
isbn = {9781713829546},
publisher = {Curran Associates Inc.},
address = {Red Hook, NY, USA},
booktitle = {Proceedings of the 34th International Conference on Neural Information Processing Systems},
articleno = {326},
numpages = {12},
location = {Vancouver, BC, Canada},
series = {NIPS '20}
}

@article{crosscode2vec2025,
  title={{CrossCode2Vec: A unified representation across source and binary functions for code similarity detection}},
  author={Yu, Gaoqing and An, Jing and Lyu, Jiuyang and Huang, Wei and Fan, Wenqing and Cheng, Yixuan and Sui, Aina},
  journal={Neurocomputing},
  doi = {10.1016/j.neucom.2024.129238},
  volume={620},
  pages={129238},
  year={2025},
  publisher={Elsevier}
}

@inproceedings{xu2023pem,
  title={{PEM: Representing Binary Program Semantics for Similarity Analysis via a Probabilistic Execution Model}},
  author={Xu, Xiangzhe and Xuan, Zhou and Feng, Shiwei and Cheng, Siyuan and Ye, Yapeng and Shi, Qingkai and Tao, Guanhong and Yu, Le and Zhang, Zhuo and Zhang, Xiangyu},
  booktitle={Proceedings of the 31st ACM Joint European Software Engineering Conference and Symposium on the Foundations of Software Engineering(FSE)},
  pages={401--412},
  doi = {10.1145/3611643.3616301},
  year={2023}
}

@article{jia20231,
  title={{1-to-1 or 1-to-n? Investigating the Effect of Function Inlining on Binary Similarity Analysis}},
  author={Jia, Ang and Fan, Ming and Jin, Wuxia and Xu, Xi and Zhou, Zhaohui and Tang, Qiyi and Nie, Sen and Wu, Shi and Liu, Ting},
  journal={ACM Transactions on Software Engineering and Methodology},
  volume={32},
  number={4},
  pages={1--26},
  year={2023},
  issue_date = {July 2023},
  publisher = {Association for Computing Machinery},
  address = {New York, NY, USA},
  issn = {1049-331X},
  url = {https://doi.org/10.1145/3561385},
  doi = {10.1145/3561385},
}

@inproceedings{jia2024codeextract,
author = {Jia, Lichen and Wu, Chenggang and Zhang, Peihua and Wang, Zhe},
title = {CodeExtract: Enhancing Binary Code Similarity Detection with Code Extraction Techniques},
year = {2024},
isbn = {9798400706165},
publisher = {Association for Computing Machinery},
address = {New York, NY, USA},
url = {https://doi.org/10.1145/3652032.3657572},
doi = {10.1145/3652032.3657572},
booktitle = {Proceedings of the 25th ACM SIGPLAN/SIGBED International Conference on Languages, Compilers, and Tools for Embedded Systems},
pages = {143–154},
numpages = {12},
location = {Copenhagen, Denmark},
series = {LCTES 2024}
}

@inproceedings{zhan2024ps3,
  title={{Ps3: Precise patch presence test based on semantic symbolic signature}},
  author={Zhan, Qi and Hu, Xing and Li, Zhiyang and Xia, Xin and Lo, David and Li, Shanping},
  booktitle={Proceedings of the IEEE/ACM 46th International Conference on Software Engineering},
  doi = {10.1145/3597503.3639134},
  pages={1--12},
  year={2024}
}

@inproceedings{ming2017binsim,
  title={{$\{$BinSim$\}$: Trace-based semantic binary diffing via system call sliced segment equivalence checking}},
  author={Ming, Jiang and Xu, Dongpeng and Jiang, Yufei and Wu, Dinghao},
  booktitle = {26th USENIX Security Symposium (USENIX Security 17)},
  year = {2017},
  isbn = {978-1-931971-40-9},
  address = {Vancouver, BC},
  pages = {253--270}
}

\appendix

\end{document}